\documentclass[%
 reprint,
 amsmath,amssymb,
 aps,
]{revtex4-2}

\usepackage{graphicx}
\usepackage{dcolumn}
\usepackage{bm}
\usepackage{xcolor}
\usepackage{hyperref}
\usepackage{siunitx, sidecap, subcaption, enumitem}
\usepackage{ulem}

\DeclareMathOperator{\Var}{\mathrm{Var}}

\begin{document}

\title{Contact map dependence of a T cell receptor binding repertoire}

\author{Kevin Ng Chau}
\affiliation{Physics Department, Northeastern University, Boston MA 02115}
\author{Jason T. George}%
\affiliation{Center for Theoretical Biological Physics, Rice Univ., Houston, TX 77005}
\author{Xingcheng Lin}
\affiliation{Dept. of Chemistry, Mass. Inst. of Technology, Cambridge, MA 02139
}%
\author{Jos\'e N. Onuchic}%
\affiliation{Center for Theoretical Biological Physics, Rice Univ., Houston, TX 77005}

\author{Herbert Levine}
\affiliation{
Center for Theoretical Biological Physics and
Depts. of Physics and Bioengineering \\  Northeastern University, Boston MA 02115.}

\date{\today}

\begin{abstract}
The T cell arm of the adaptive immune system provides the host protection against unknown pathogens by discriminating between host and foreign material.  This discriminatory capability is achieved by the creation of a repertoire of cells each carrying a T cell receptor (TCR) specific to non-self antigens displayed as peptides bound to the major histocompatibility complex (pMHC). The understanding of the dynamics of the adaptive immune system at a repertoire level is complex, due to both the nuanced interaction of a TCR-pMHC pair and to the number of different possible TCR-pMHC pairings, making computationally exact solutions currently unfeasible. To gain some insight into this problem, we study an affinity-based model for TCR-pMHC binding in which a crystal structure is used to generate a distance-based contact map that weights the pairwise amino acid interactions. We find that the TCR-pMHC binding energy distribution strongly depends both on the number of contacts and the repeat structure allowed by the topology of the contact map of choice; this in turn influences T cell recognition probability during negative selection, with higher variances leading to higher survival probabilities. In addition, we quantify the degree to which neoantigens with mutations in sites with higher contacts are recognized at a higher rate.

{\color{magenta} }
\end{abstract}

\maketitle


\section{Introduction}\label{sec:intro}

One of the major components of the human immune system consists of a large repertoire of T lymphocytes (or T cells). Each T cell carries a particular T cell receptor (TCR) capable of binding to a specific antigen in the form of a peptide (p) displayed by major histocompatibility complex (MHC) molecules (shortened as pMHC) on the surface of host cells \cite{Ding2012, Robinson2016, Schumacher2015, Verdegaal2016}. The activation of the T cell response depends on the strength \cite{Das2015}, and possibly kinetics \cite{Francois2016}, of this TCR-pMHC binding \cite{Alam1996, Krogsgaard2005}. A typical repertoire of a healthy individual consists of  $\sim10^7$ distinct clonotypes, each with a unique TCR \cite{Arstila1999}. A growing body of research has been focused on understanding the systems-level interactions between the T cell repertoire and its recognition of peptide landscapes indicating foreign or cancer threats.

A critical feature of a properly functioning immune system is its ability to discriminate healthy cells of the host from those infected by pathogens, reacting to the latter ones while tolerating the former ones. In order to achieve the aforementioned discrimination, T cells must survive a rigorous selection process in the thymus before being released into the bloodstream. The first step in this process, called positive selection, ensures that TCRs in thymocytes (developing T cells) can adequately interface with pMHCs. Positive selection occurs in the thymic cortex, where cortical epithelial cells present self-peptides to thymocytes.  As long as a thymocyte is able to interface with some presented pMHC, it receives a survival signal and migrates inward to the thymic medulla. This step ensures that the thymocyte has a properly functioning TCR, a rare event as only 5\% of thymocytes survive this step. In the inner medulla, they encounter thymic medullary epithelial cells. Here, surviving immature T cells are again presented with a diverse collection of $\sim10^4$ self peptides \cite{DeBoer1993, Yates2014} representing a variety of organ types. T cells binding too strongly to any self peptide die off in a process known as negative selection \cite{Detours1999, Klein2014}.

As already pointed out, a key ingredient in the aforementioned process as well as in any subsequent recognition of a foreign antigen by a T cell is the molecular interaction  of the TCR and the pMHC molecules.  Crystal structures of TCR bound to pMHC show that the interface of the TCR-pMHC interaction is complex, with TCR complementarity determining regions 1 and 2 (CDR1 and CDR2, respectively) primarily binding to the MHC molecule, whereas the CDR3 complex mainly contacts the peptide in the MHC's cleft \cite{Lanzarotti2011, Newell2011}. The CDR3 complex is comprised of two loops, CDR3$\alpha$ and CDR3$\beta$; Baker et al. showed these loops can exhibit spatial and molecular flexibility during the TCR-pMHC binding process \cite{Baker2012}; moreover, the same TCR can bind to different pMHCs \cite{Colf2007}, for example to a pMHC with point-mutated peptide \cite{Newell2011}. This can involve subtle changes in the CDR3 complexes' spatial conformation. It is clear then that the intricacies of the TCR binding to the pMHC as a dynamic process remain as yet to be fully understood.

In lieu of a complete first-principles understanding, several groups have pioneered the idea of employing relatively simple models so as to get a sense of how negative selection affects the T cell repertoire. In the original set of models, TCRs and peptides were represented as strings of amino acids (AAs) which interacted in a manner that did not incorporate any structural information. In one such set of models, each AA in the pMHC binding pocket interacted with and only with the complementary AA in the TCR CDR3 complex. This interaction was described by either one or a set of 20x20 matrices \cite{Kosmrlj2008, Kosmrlj2009, Chakraborty2010, George2017, Wortel2020}. These works indeed have provided a framework for describing how selection shapes the discrimination ability of the T cell repertoire, and have been applied to understanding HIV control \cite{Kosmrlj2010} and for assessing the detectability of cancer neoantigens \cite{George2017}. In a more recent study, Chen et al. \cite{Chen2018} introduced nonuniform interaction profiles that translated into some AAs in the TCRs having a more pronounced effect in pMHC recognition, but did not consider how these non-uniformities could vary between TCRs, as shown by existing crystal structures.

In this paper, we introduce the idea of a crystal-structure dependent contact map that weights the binding energies based on the distance separating the residues on the AAs. A contact map can be thought of as a specific template for a class of TCR-pMHC interactions, which then will yield an actual binding energy once we specify the specific AA strings on the two molecules. To focus attention on the role of the contact map, we use a simple random energy model which assigns a fixed random energy to each of the possible AA pairs. Our model, described in detail below, can be thought of a more realistic version of the the Random Interaction Between Cell Receptor and Epitope (RICE) model \cite{George2017}, in which contact map effects were simply assumed to decorrelate pair energies at different sites along a uniform binding surface. 

The paper is structured as follows. In section \ref{sec:model}, we present the model description along with how crystal-structure dependent contact maps are created and also discuss the choice of energy matrix in the model. In Section \ref{sec:PDF}, we analyze how the variance of the TCR-pMHC binding energy PDF is impacted by the choice of contact map, including the roles of the total number of contacts and the topology of the contact map. We then present two applications of the model that are affected by the choice of contact map: in Section \ref{sec:NegSelSP} we focus on the negative selection recognition probability, and in Section \ref{sec:PMRecProb} we discuss the point-mutant recognition probability by T cells that have survived negative selection. We present our closing remarks in Section \ref{sec:conclusions}.

\begin{figure*}
    \centering
    \begin{subfigure}[b]{.34\textwidth}
        \includegraphics[width=\textwidth]{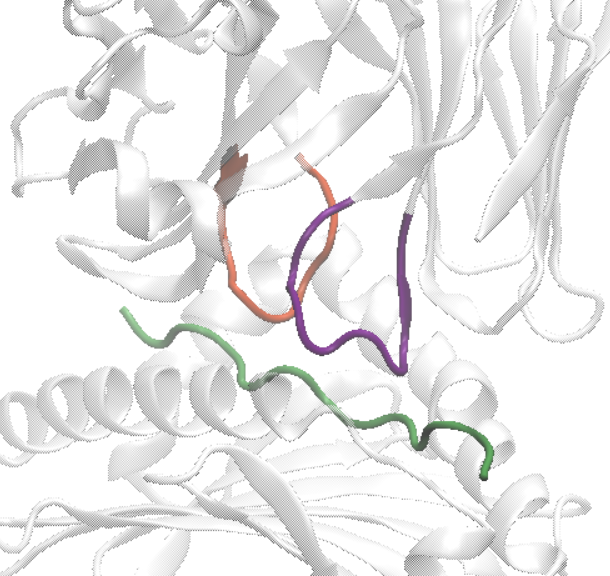}
        \caption{}
        \label{fig:Crystal}
    \end{subfigure}
    \begin{subfigure}[b]{.65\textwidth}
        \includegraphics[width=\textwidth]{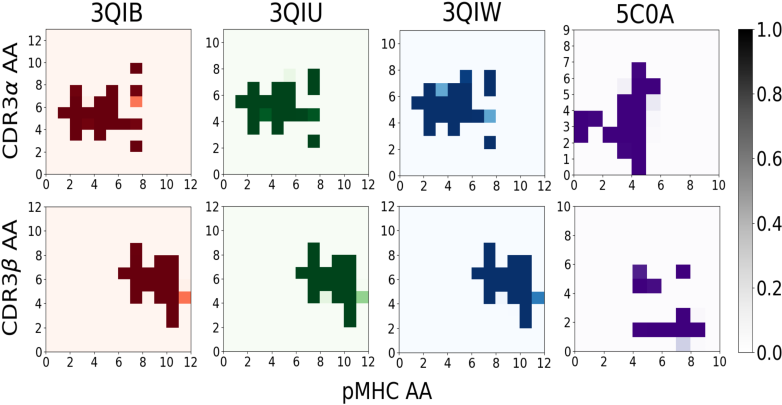}
        \caption{}
        \label{fig:CMs}
    \end{subfigure}
    \caption{\small The TCR-pMHC interface and contact maps. Panel \ref{fig:Crystal} shows the CDR3-pMHC interface in the crystal structure of the 2B4 TCR binding to the MCC/I-E\textsuperscript{k} complex (PDB ID 3QIB); with the antigen MCC highlighted in green, the CDR3$\alpha$ loop in purple, and the CDR3$\beta$ loop in orange. Panel \ref{fig:CMs} shows eight contact maps estimated from four crystal structures, contact maps of the CDR3$\alpha$-pMHC (CDR3$\beta$-pMHC) interfaces in the top (bottom) row; 3QIB, 3QIU, and 3QIW are MHC class-II restricted, whereas 5C0A is MHC class-I restricted.}
    \label{fig:ContactMaps}
\end{figure*}
\section{Contact Map Based Random Energy Model}\label{sec:model}

Our goal is to analyze a model of negative selection in which the TCR-pMHC interaction exhibits antigen-specificity of T cells dependent both on the AA occurrence and on the spatial conformation of TCR and pMHC, while retaining enough simplicity so that it can be studied analytically and with feasible computations. We represent a TCR $t$ via its CDR3 loops in form of a sequence of $k_t$ AAs, $t = \{ t(i) \}_{i=1}^{k_t}$, and a pMHC $q$ as a sequence of $k_q$ AAs, $q = \{ q(j) \}_{j=1}^{k_q}$. A symmetric energy coefficient matrix of size $20 \times 20$, $\mathbb{E} = (E_{nm})$, has entries $E_{nm}$ that represent the pairwise binding coefficients between AAs $n$ and $m$. The binding energy contributions are then assumed to be the product of a contact map $\mathbb{W} = (W_{ij})$, containing the weights $W_{ij}$ for the interaction between $t$ and $q$ in a given structure, and the coefficient corresponding to the amino acid interaction. In detail, 
\begin{equation}\label{Etq}
    U(t, q) \,=\, U_c \;+\; \sum_{i, j} W_{ij} \cdot E_{t(i)q(j)},
\end{equation}
where $U_c$ represents the contribution of the TCR's CDR1 and CDR2 complexes interacting with the MHC molecule, as discussed in \cite{Kosmrlj2008, Kosmrlj2009, Chakraborty2010, Kosmrlj2010}.

This form of the binding energy in \eqref{Etq} explicitly separates the effects on CDR3-pMHC interaction due to spatial configuration from the effects due to the rest of the pair-dependent factors, assigning the former ones to $\mathbb{W}$ and coarsely accounting for the latter ones in $\mathbb{E}$. The particular choices for the contact map $\mathbb{W}$ will depend on the specific TCR-pMHC being used as a template. Also, this formulation does not pre-suppose any specific choice for $\mathbb{E}$.  We discuss in detail specific choices of $\mathbb{E}$ and $\mathbb{W}$ in the sections below.

\subsection{Contact maps}\label{sec:ContactMaps}


Crystal structures of TCRs bound to pMHCs show a variety of spatial configurations. Each one of these can be thought of as defining a binding template which can be used to determine the energy of a set of possible pairs. In general, we expect there to be a small number of possible templates, as a specific template would presumably be valid for a subset of all pairs; even then, we must necessarily ignore the small structural changes seen between the same TCR-pMHC systems that differ e.g. by a single AA mutation \cite{Cole2016, Newell2011, Sethi2013, Ting2020}. We expect, based on a recent computational study \cite{Lin2021}, that this approach will be reasonable if we stick to a fixed MHC allele, as structures with different alleles can look very different. We will see this directly in Fig. \ref{fig:ContactMaps} below. In the calculations reported in this paper, we typically restrict ourselves to one template.

To derive a contact map from a crystal structure, we utilize the associative memory, water mediated, structure and energy model (AWSEM) \cite{Davtyan2012}, developed in the context of protein folding. We use the position of C$_\beta$ (C$_\alpha$ in the case of glycine) atoms to characterize the position of the residues of the AAs in both the TCRs and pMHCs, and to use AWSEM's negative-sigmoid switching function as the screening weight $W_{ij}$ in computing the interaction energy
\begin{equation}\label{Wij}
    W_{ij} (r_{ij}) = \frac{1}{2} \, \left( 1 - \tanh{[\eta \cdot (r_{ij} - r_{\text{max}})]} \right).
\end{equation}
Here, $r_{ij}$ is the distance separating the residues at positions $i$ and $j$,  $r_{\text{max}}$ acts like a cutoff and is the inflection point of $W_{ij}$ after which the function vanishes rapidly for $r_{ij} > r_{\text{max}}$, and $\eta$ controls how rapidly this vanishing occurs. We use crystal structures (see Fig. \ref{fig:ContactMaps}a) of TCR bound to pMHC deposited in the Protein Data Bank to determine a list of AAs in the TCR $t$ and in the pMHC $q$, and to calculate each distance $r_{ij}$, $i = 1, \cdots, k_t$, $j = 1, \cdots, k_q$. We then compute the corresponding weights $W_{ij}$ from \eqref{Wij} and construct the contact map $\mathbb{W} = (W_{ij})$. Given that both CDR3$\alpha$ and CDR3$\beta$ loops of the TCR interface with the peptide, we construct a separate contact map for each of these CDR3-loop-pMHC interactions.

To show how the proposed screening weight given by \eqref{Wij} derives from different TCR-pMHC crystal structures, we choose $r_{\text{max}} = 9.5 \; \si{\angstrom}$ and $\eta = 1 \; \si{\angstrom}^{-1}$ and focus on four test cases. For the first three test cases we use data from Newell et al. \cite{Newell2011} who present three TCR-pMHC crystal structures; first, of the 2B4 TCR bound to the MCC/I-E\textsuperscript{k} complex (PDB ID 3QIB); second, of the 226 TCR bound to the MCC/I-E\textsuperscript{k} complex (PDB ID 3QIU), and; third, of the 226 TCR bound to the MCC-p5E/I-E\textsuperscript{k} complex (PDB ID 3QIW). For the fourth case, we follow Cole et al. \cite{Cole2016} who studied the 1E6 TCR bound to HLA-A02 carrying MVW peptide (PDB ID 5C0A). For simplicity, we will refer to specific crystal structures by their PDB IDs unless further details need to be more precisely mentioned about the TCR or the pMHC. Note that 3QIB and 3QIU represent different TCRs bound to the same pMHC complex, whereas 3QIU and 3QIW represent the same TCR bound to two pMHCs that differ by a single AA mutation in the peptide sequence. In addition, 3QIB, 3QIU, and 3QIW share the same mouse MHC class-II restriction and indeed the same I-E$^\textup{k}$ MHC-II allele, whereas the 5C0A TCR-pMHC system is presented on the human HLA A$^*$02 MHC class-I allele.

As defined here, contact maps are sensitive to the choice of distance cutoff. Clearly, the number of contacts in a contact map for a given crystal structure increases with increasing $r_{\text{max}}$ values. The contact map of 3QIB's CDR3$\alpha$-pMHC interface is plotted at four different $r_{\text{max}}$ values, from 6.5 to 9.5 \si{\angstrom} in 1 \si{\angstrom} increments, while keeping $\eta = 1 \, \si{\per\angstrom}$ fixed  (see SI Fig. S1). The contact profile gradually forms with ever-increasing number of contacts from about 5 AA pairs in contact at $r_{\text{max}} = 6.5 \, \si{\angstrom}$, to about 22 AA pairs in contact at $r_{\text{max}} = 9.5 \, \si{\angstrom}$. For the remainder of this paper, all contact maps are calculated with $r_{\text{max}} = 9.5 \, \si{\angstrom}$ and $\eta = 1 \, \si{\per\angstrom}$.  

The contact maps in figure \ref{fig:CMs} correspond to CDR3$\alpha$-pMHC interfaces (top row) and CDR3$\beta$-pMHC interfaces (bottom row) from crystal structures 3QIB, 3QIU, 3QIW, and 5C0A. The contact profiles of CDR3$\alpha$-pMHC are different from the CDR3$\beta$-pMHC contact profiles, as these parts of the TCR contact different residues on the displayed peptide. The contact maps consistently represent the physical proximity of a particular CDR3 loop to a specific portion of the pMHC, as can be seen in 3QIB's crystal structure shown in figure \ref{fig:Crystal}, wherein the CDR3$\alpha$ loops primarily contact AAs 2-8, whereas CDR3$\beta$ loops primarily contact AAs 7-12. The detailed differences among the first three contact maps do capture slight changes in position-dependent interfacing, even when comparing contact maps for the same TCR bound to two pMHCs diverging by peptide single-AA mutation. Different weights of, for example, position pairs $(i,j)$ = $(4, 4)$, $(4, 8)$, $(6, 4)$ and $(7, 6)$ are observed when comparing contact maps of 3QIU and 3QIW in figure \ref{fig:CMs} (coordinates in AA pairs are labeled as $(i, j)$ for $t(i)$ and $q(j)$). But, clearly from a more coarse-grained perspectives, these three can be considered to fall within one template. Conversely, the fourth map is very different, as should be expected because it is based on a different MHC molecule. Our conclusion is that we can use a single map for a class of possible parings and thereby learn about a significant set of contributors to the T cell repertoire. We include more contact maps from other crystal structures in the SI to further support our findings (Figs. S2-4).
In the remainder of this paper, we will explore the segment of the repertoire that depends on one template and its corresponding contact map, and determine how the features of that map affect repertoire properties.

\subsection{Energy matrix}\label{sec:EnergyMatrix}

As discussed above, we propose for the recognition of an antigen by a T cell an affinity-based criterion in which the TCR-pMHC binding energy $U(t, q)$ given in \eqref{Etq} equates to recognition (evasion) if $U(t, q)$ is above (below) a particular energy threshold $U_n$. Thus, we need to specify a symmetric energy coefficient matrix $\mathbb{E} = (E_{nm})$. The first example of matrix choice was one based primarily on hydrophobicity, as developed by Miyazawa-Jerningan (MJ) \cite{Miyazawa1985} and used in studies of thymic selection \cite{Kosmrlj2008, Chen2018}. More recent efforts have focused on developing immune-specific energy matrices \cite{Woelke2011}.  A recent study \cite{Lin2021}  used machine learning to derive the optimal matrix separating strong from weak binders within a single contact map template; this optimization approach would lead to a different such matrix for each assumed template. Here, our interest is in the role of the contact map and so we have opted for the expedient choice of a random model where all matrix elements are chosen to be  independent, mean-zero, unit-variance normally distributed random variables, $E_{mn} \sim \mathcal{N} (\mu = 0, \sigma^2 = 1)$. Note the assumption that the $n$-$m$ interaction coefficient has the same value independently of the AAs' location in the TCR or the pMHC sequences. Thus, our model is distinct from the RICE approach \cite{George2017} which assumed that the spatial location of the amino acid directly affected the energy coefficient.

\begin{figure*}
    \begin{centering}
    \begin{subfigure}[b]{0.48\textwidth}
        \includegraphics[width=\textwidth]{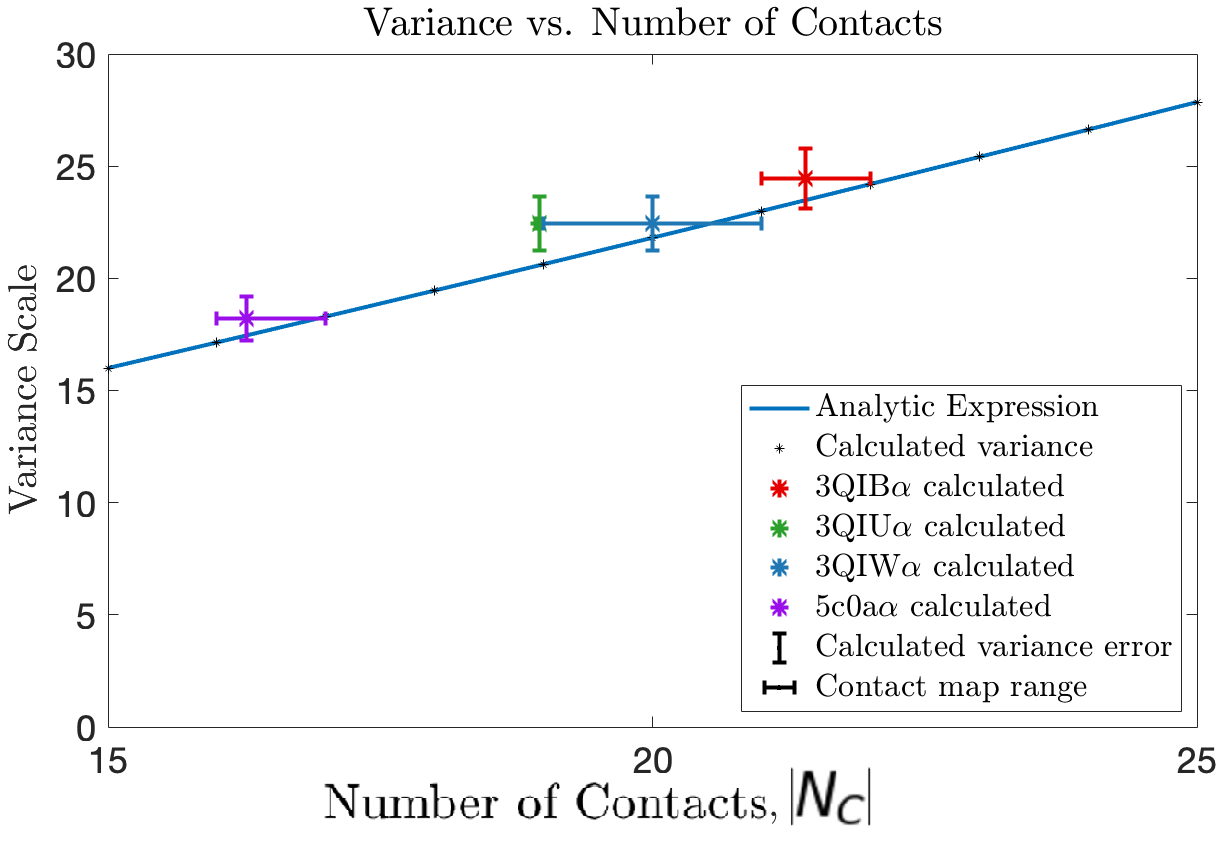}
        \caption{}
        \label{fig:VarvsR}
    \end{subfigure}    
    \begin{subfigure}[b]{0.45\textwidth}
        \includegraphics[width=\textwidth]{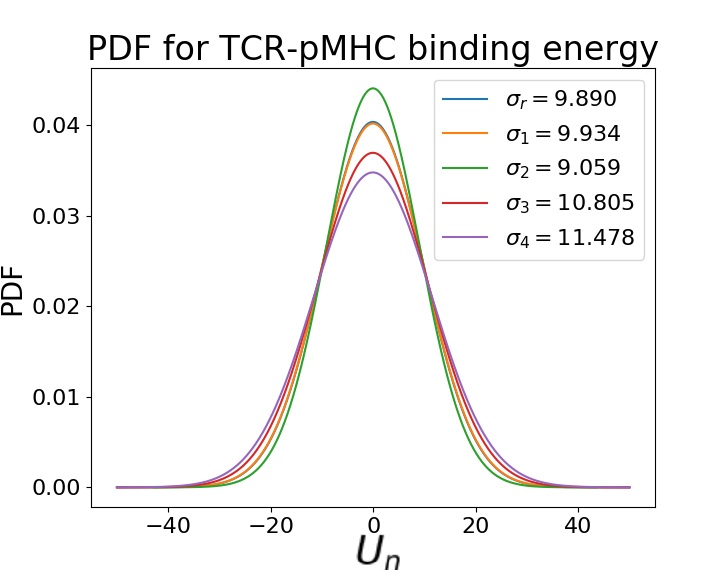}
        \caption{}
        \label{fig:VarPartitions}
    \end{subfigure}
    \end{centering}
    \caption{The variance of the TCR-pMHC binding energy distribution depends on the total number of contacts and on the repeat structure allowed by the topology in the contact map. Panel \ref{fig:VarvsR} shows binding energy $U(t, q)$ variance scaling with the number of contacts $|N_C|$;  calculated variances with their variance (vertical error bars) were plotted as a function of total contacts $|N_C|$ in the contact map. Horizontal error bars represent the range of threshold used for determining each contact map (lower estimates corresponding to counting contacts $>$ 0.9, and upper estimates corresponding to including contacts $>$ 0.1). Panel \ref{fig:VarPartitions} shows the binding energy PDFs and corresponding simulated standard deviations ($\sigma_r$, $\sigma_1$, etc) for pMHC repertoires of randomly chosen AA sequences (blue) and with all TCRs constrained to the same repeated AAs motif; repertoires constrained to each of the four most likely pMHC repeat motifs are shown with different colors and are labelled in decreasing order of likelihood. In simulations $\sigma^2 = 1$.}
    \label{fig:Var}
\end{figure*}

\section{Distribution of TCR-pMHC binding energy}\label{sec:PDF}

The TCR-pMHC binding energy $U(t, q)$ is the indicator of the affinity between a T cell and an antigen. When assuming the pairwise AAs interaction energies to be independent Gaussian random variables, $U(t, q)$ in \eqref{Etq} becomes a weighted sum of these variables with weights given by the contact map $\mathbb{W}$. Hence, $U(t, q)$ is also a normally distributed random variable, and since its mean is automatically zero,  knowledge of the variance $\sigma^2_{tq}$ of its PDF allows us to fully characterize how $U(t, q)$ varies as we vary the particular realization of $\mathbb{E}$. The contact-map dependence of $U(t, q)$ has a two-fold impact on the variance of its PDF when compared to the case of the addition of equal variance random variables (as in the RICE approach from \cite{George2017}). On one hand, the total number of non-vanishing contacts $W_{ij}$ given by the contact map directly determines the number of random energies $E_{ij}$ contributing to $U(t, q)$, thus increasing $\sigma^2_{tq}$ as the number of non-vanishing $W_{ij}$'s increases. On the other hand, the particular repeat structure of AAs in the TCR sequence and in the pMHC sequence also influences $\sigma^2_{tq}$, as a particular pair of AAs that appears multiple times in the energy summation gives rise to a variance increase.  In this section we explore how the variance of the PDF of $U(t, q)$ depends on the two aforementioned factors.

Before proceeding, we must discuss various statistical ensembles of interest here. So far, we have focused on varying the coefficient matrix, thus generating an ensemble values for each specific $t,q$. However, we imagine that the biophysical problem is defined by a fixed $\mathbb{E}$, which may be chosen (as done here) in a random fashion but, as mentioned above, may be learned from the data as done in other work \cite{Lin2021}. Thus, we are actually interested in the distribution of binding energies as we vary either the peptide (fixing the TCR), the TCR (fixing the peptide) or both, as these are what is necessary to determine the effects of negative selection. To see how to determine these distributions, we return to the basic equation
\begin{equation}\label{EnergyLoop}
    U(t, q)  = \sum_i^{k_t}  \sum_j^{k_q} W_{ij} \cdot E_{t(i)q(j)},
\end{equation}
where we have limited ourselves to one class of MHC molecule and hence $U_c$ becomes an irrelevant constant. Also, we will assume for the purpose of our analysis that $W_{ij}$ is either 0 or 1; this is true for all but a very small number of possible pairs. Finally, we will assume take the distribution over AA to be uniform, although it might be useful in future work to use the known AA distribution in the human proteome. With these number of assumptions, the mean value of $U (t,q)$ sampled over peptide sequence and/or TCR sequence constrained to have no repeats is just the sample mean of drawing a number of values from a mean zero, variance $\sigma^2$ Gaussian distribution.    This number is very much peaked around zero. Similarly, the mean value of $U^2$ will be strongly peaked around the variance times the contact number $N_c$. Perhaps not surprisingly, these are the same answers we get when averaging over $\mathbb{E}$; in other words, as long as we average over sufficient numbers of sequence choices, the results for all choices of coefficient matrices are the same; see the SI (Section S8) for a more complete discussion.

Let us now extend this analysis to the more general case. We introduce the following notation: A pair repeat structure is denoted as $C_p = (l_1^{r_1}, l_2^{r_2}, \cdots, l_N^{r_N})$, with $\sum r_i \cdot l_i = N_C$, where $l_i$ denotes the number of times an amino acid pair is repeated in different contacts and $r_i$ denotes how many such $l_i$ repetitions there are. For example, for a total of 20 contacts, if there are three contacts with the same AA pair and two set of two contacts with the same AA pair, this would be denoted as $C_p=(3,2^2,1^{13})$. An extension of the previous argument allows us to determine the most likely value of the mean energy and its variance, averaged over all possible peptide and TCR sequences that do not change the class. The mean is still zero and the variance now becomes
\begin{equation}\label{EqVariance}
    \textup{Var}(C_p) = \sigma^2 \sum r_i l_i ^2.
\end{equation}
Again, this is exactly the same as the result obtained when averaging over energy coefficient matrices. A more precise version of this correspondence is presented in the SI (S5 and S6). If one wants to find the total variance, we have to average over different choices of $C$ weighted by their respective probabilities of occurrence given the assumed uniform distribution of residue choice.

\subsection{Variance scales with the number of contacts}

It is clear from the previous analysis that the variance in the binding energy distribution increases with $N_C$ the total number of contacts. It is easy to see from the above that there are bounds on the total variance
\begin{equation}\label{EqVarianceBounds}
   \sigma^2  N_C \leq \Var U   \leq  \sigma^2 N_C^2 .
\end{equation}
The lower bound comes from the case where all pairs are distinct whereas the upper bound arises from assuming that all contacts are the same AA pair, i.e. $C=(N_C)$. From the size of the AA alphabet $|\mathcal{A}|$, the total number of AA pairs (irrespective of ordering) is $M = \binom{|\mathcal{A}| + 1}{2}$. Now, we have just seen that the precise value of the variance depends on the exact repeat structure of the peptide ($q$) and TCR ($t$) AA sequences, together with the contact map.  In the case where we wish to obtain the variance of the PDF obtained by varying both $t$ and $q$, we can obtain a useful approximation of this variance by ignoring the exact configuration of $\mathbb{W}$ and instead simply counting the number of times each of the $M$ AA pairs is selected with equal probability, where there are $N_c$ total opportunities. In this case, the number of times each AA pair is realized follows a multinomial distribution, and the variance can be calculated from the second moment of this distribution as
\begin{equation}\label{EqVarvsR}
    \textup{Var}\left(U(t,q) | \mathbb{W}\right) \approx \frac{1}{M} N_C^2 + \left( 1 - \frac{1}{M} \right) N_C.  
\end{equation}
See the SI (Sections S5 and S6) for a detailed derivation. In figure \ref{fig:VarvsR} the variances computed by simulation for the CDR3$\alpha$-pMHC interfaces of 3QIB, 3QIU, 3QIW, and 5C0A (top row of Fig. \ref{fig:CMs}) are presented along with the predicted variance from \eqref{EqVarvsR}. As we can see, this approximation captures the basic dependence on the total number of contacts. In the SI (Fig. S5), we provide further evidence for this result by considering the effects of varying the cutoff used in the definition of the contact matrix. 

\subsection{Variance depends on the repeat structures of the TCR and pMHC AA sequences}

If we are looking for the distribution of energies for a fixed TCR sequence, there is no simple formula that can encompass the dependence of the variance on the exact TCR sequence and on the exact contact map. As already mentioned, we have to find the variance for different possible repeat structures and then weight them appropriately by their occurrence probability. Specifically,
\begin{equation}\label{VarRSGen}
    \sigma_{t}^2 \,=\, \sum_{n=1}^{N_R} p_n \sigma_n^2.
\end{equation}
Where $N_R$ is the total number of different possible structures.

We would like to work out a specific and relatively simple example to illustrate how this works. To simplify the analysis, we focus on the 3QIB CDR3$\alpha$-pMHC contact map $\mathbb{W}_{\text{3QIB}}^{\alpha}$ in figure \ref{fig:CMs} (top left), and assume that the TCR is a constant sequence of a single repeated AA $t = (t_1,t_1,t_1...)$. Note that this makes labelling of repeat motifs dependent on the pMHC's primary sequence only. In $\mathbb{W}_{\text{3QIB}}^{\alpha}$, only 7 AAs in $t$ and 7 AAs in $q$ make significant contacts, so the effective lengths are $k_t = k_q = 7$. 

We will break down the problem of computing the terms in this sum as follows: We will first focus on the probable configurations of the peptide by itself and consider how the different sites are chosen.  Drawn from a $|\mathcal{A}| = 20$ AA alphabet, there are $N = 15$ different repeat configurations of length 7; when randomly generating AA sequences, the four most likely repeat configurations $C_{q,1} = (2, 1^5)$, $C_{q,2} = (1^7)$, $C_{q,3} = (2^2, 1^3)$, and $C_{q,4} = (3, 1^4)$ (in the section above, $C$ is the repeat structure of the TCR-pMHC pairing, whereas $C_{q,n}$ ($n = 1, \cdots, N$) here indicate the repeat structure only of the pMHC), cover about $p_c = 96.66 \%$ of the AA sequence space. A complete breakdown of these probabilities can be found in  SI Table 1. We thus truncate the sum in \eqref{VarRSGen} to the pairings that can be obtained from these leading order structures. 

Now, each peptide configuration can give rise to a set of different possible pairing structures, depending on the specific non-vanishing elements of the contact matrix. These then need to be averaged together (with proper weighting). This somewhat complicated calculation is presented in the SI (section S6) and is carried out by using the self-averaging property to allow for computing the average over different realizations of the energy coefficient matrix; no rounding to 0 or 1 for the values $W_{ij}$ is made in this calculation and the results to follow. Finally, we obtain $\sigma_{t} (p_c) = 9.7833 \sigma$, and extrapolating this value to approximate the full analytical value in \eqref{VarRSGen}, we get
\begin{equation*}
    \sigma_{t} \approx \sqrt{\dfrac{1}{p_c}} \cdot \sigma_{t} (p_c) = 9.95 \sigma.
\end{equation*}
This estimation has relative error of $0.6 \%$ as compared to the simulated value of the standard deviation, the blue plot in figure \ref{fig:VarPartitions}. The simulated PDFs related to the four most likely repeat structures are also shown in figure \ref{fig:VarPartitions}.

It is worth noting that in \eqref{VarRSGen} the contributions of higher values of variances are dominated by the even faster vanishing of the corresponding probabilities. For reference, the standard deviation for this contact map ranges from $\sigma_2 = 9.0761$ for $C_{q,2} = (1^7)$ to $\sigma_{15} = 21.4090$ for $C_{q,15} = (7)$; whereas the probabilities are $p_2 = 30.52 \%$ and $p_{15} = 1.56 \times 10^{-6} \%$, respectively.

\section{Negative selection recognition probability}\label{sec:NegSelSP}

\begin{figure}
    \includegraphics[width=0.45\textwidth]{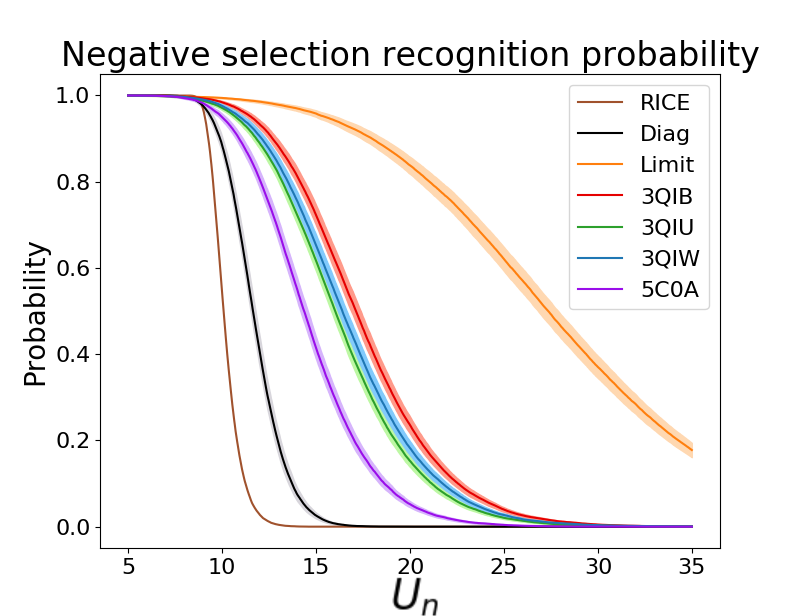}
    \caption{\small Negative selection recognition probability as a function of the survival energy threshold for T cells auditioning for negative selection. All curves involving the use of contact maps are generated from simulations sharing the same parameters apart from the contact maps. The prediction of the RICE model (brown), the identity matrix giving a diagonal contact map case (black) and the limiting case where all AAs in the CDR3 loop interact with all AAs in the pMHC (yellow) are included for comparison. Plots are averaged over the different random energy matrices in use, and shaded areas indicate the corresponding standard error of the mean.}
    \label{fig:NegSelRecProb}
\end{figure}

Negative selection trains the na\"{\i}ve T cell repertoire to avoid host cells by eliminating T cells that bind too strongly to any of the self-peptides. We now wish to consider the effects on the post-selection repertoire due to incorporating crystal-structure motivated contact maps into the negative selection process.

We focus on determining the negative selection recognition probability as a function of the energy survival threshold $U_n$. For a T cell to survive negative selection, it must not bind strongly i.e. $U < U_n$  to any of the self-selecting pMHCs it encounters during selection. This is described by the probability that the maximum of the TCR-pMHC binding energies, $\max \{U(t, q_i)\}_{i=1}^{N_q}$, resulting from a T cell $t$ undergoing negative selection against a repertoire, $\mathcal{Q} = \{ q_i \}_{i=1}^{N_q}$ of $N_q$ self-pMHCs, is below the threshold $U_n$\cite{George2017}. This recognition probability is thus a monotonically decreasing function that gradually transitions from 1 to 0 with ever increasing values of $U_n$. For a fixed TCR, the scale of the transition correlates with a typical value of $\sigma^2_{t}$. Averaging this over different TCRs will give rise to a width that strongly correlates with the number of contacts, as suggested by the phenomenological relationship given above and verified in the SI.

We simulate negative selection for various CDR3-pMHC interfaces (contact maps), using fixed randomly generated TCR and pMHC repertoires and 16 zero-mean, unit-variance randomly generated energy matrices $\mathbb{E}$. In figure \ref{fig:NegSelRecProb}, we show the recognition probability averaged over energy matrices $\mathbb{E}$ for seven different simulations, four of them using contact maps 3QIB, 3QIU, 3QIW, and 5C0A; along with a $7 \times 7$ identity-matrix contact map case, as well as the original RICE model, and a $7 \times 7$ contact map with all unit entries case simulating the scenario where all AAs in $t$ are interacting with all AAs in $q$. At a given $U_n$, the recognition probability is higher for those contact maps with higher $\sigma^2$, giving a higher probability for a pair of $t$ and $q$ to bind strongly enough and thus for $t$ to face deletion. Interestingly, the data in the figure show directly that, similar to what we argued earlier, the recognition probability curve for a single realization is quite accurately given by the average over energy matrices.

\begin{figure*}
    \centering
    \begin{subfigure}[b]{0.45\textwidth}
        \includegraphics[width=\textwidth]{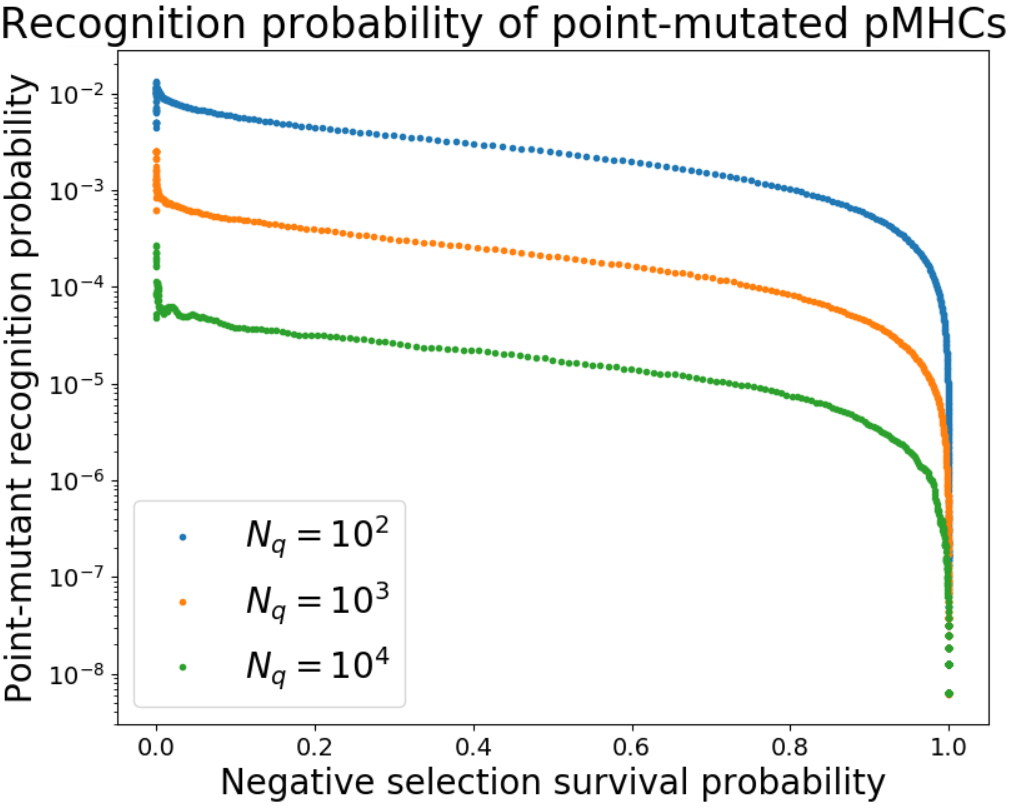}
        \caption{}
        \label{fig:PMRP-Nq}
    \end{subfigure}
    \begin{subfigure}[b]{0.45\textwidth}
        \includegraphics[width=1.1\textwidth]{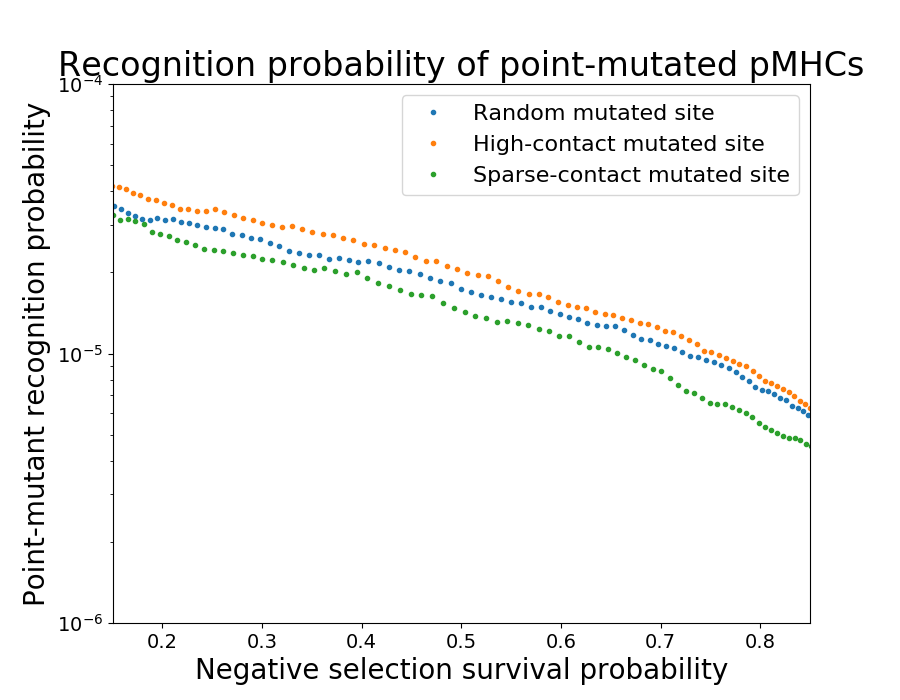}
        \caption{}
        \label{fig:PMRP-Contacts}
    \end{subfigure}
    \caption{Recognition probability of point-mutated peptides by T cells that have undergone negative selection. In panel \ref{fig:PMRP-Nq}, the point-mutant recognition probability from simulations is plotted for T cells that have received negative selection against self-peptide repertoires of three different sizes, $N_q = \{  10^2, 10^3, 10^4\}$. Panel \ref{fig:PMRP-Contacts} shows the point-mutant recognition probability from simulations that changed the site of the mutated AA; for the CDR3$\alpha$-pMHC interface of 3QIB (top left panel of Fig \ref{fig:CMs}) in use, pMHC-AAs in high-contact sites are in contact with 5 TCR-AAs, whereas pMHC-AAs in sparse-contact sites are in contact with only 1 TCR-AA; and when picking random sites to mutate, the number of peptide-AAs that a given TCR-AA can contact ranges from 1 to 5.}
    \label{fig:PMRPs}
\end{figure*}

\section{Recognition probability of point-mutated antigens by negatively-selected T cells}\label{sec:PMRecProb}

One of the motivations to model negative selection is to understand how the rejection of T cells that detect self-peptides negatively impacts the chances that T cells can detect tumor neo-antigens; after all, these neo-antigens are typically just one mutated amino acid away from a self-peptide sequence. We therefore turn to the probability that a T cell ($t$) that has survived negative selection is able to recognize an antigen ($\tilde{q}$) whose primary sequence differs by only one AA from a self-peptide ($q$) included in the negative-selecting repertoire ($\mathcal{Q}$). We call such antigen a point-mutant. In general, this probability for  fixed T cell is defined via
\begin{equation}\label{PMRecProbGen}
    \tilde{D}_t (N_q) = \mathbb{P} \left[ U(t, \tilde{q}) \geq U_n | \max\{U(t, \mathcal{Q}) \} < U_n \right],
\end{equation}
where we have averaged over all possible point-mutants with non-trivial contacts. Here $\mathcal{Q}$ denotes the selecting repertoire of $N_q$ peptides, one of which is $q$. In the limiting case where $t$ has not undergone negatively selection ($N_q = 0$), equation \eqref{PMRecProbGen} reduces to the recognition probability of a randomly generated antigen. Another extreme case corresponds to $t$ negatively trained only on $q$ ($N_q = 1$) where the point mutant position has $k$ contacts, resulting in
\begin{align}\label{PMRecProb1}
 D_t (1)   &=1-F_R(U_N)^{-1}\bigg[\int_{\mathbb{R}} F_{R-k}(U_n-x)F_k(x)f_k(x)dx \nonumber \\
                    &+ \int_\mathbb{R} \int_{[x,\infty)}F_{R-k}(U_n-\tilde{x})f_k(\tilde{x})f_k(x)d\tilde{x}dx\bigg],
\end{align}
where $F_k(x)$ and $f_k(x)$ denote the distribution function and density function of mean-zero normal random variables with variance $\sigma^2k$ (see SI Section S7 for a full derivation). We expect that for relatively small $N_q$, it is unlikely that any of the peptides in the training set will be close enough to $q$ or $\tilde{q}$ to help distinguish the two binding energies; hence $\tilde{p}_{1}$ should be a reasonable approximation to $D_t$. This agreement should decrease as $N_q$ increases. The accuracy of this approximation is explored in SI Fig. S8. 

More generally, we ran a set of simulations with varying sizes $N_q = \{ 10^2, 10^3, 10^4 \}$ to assess the detection of $\tilde{q}$ by a T cell trained to evade $q$.   We used the CDR3$\alpha$-pMHC interface of 3QIB (top left of Fig. \ref{fig:CMs}) as contact map for the simulations for simplicity. Figure \ref{fig:PMRP-Nq} shows the simulated point-mutant recognition probabilities as a function of T cell negative selection survival probability at three different sizes of the selecting repertoire. At lower (resp. higher) values of negative selection survival probability, i.e. when negative selection is more (resp. less) stringent during T cell maturation, a mature T cell's sense of an antigen resembling self-antigens is rather strict (resp. lenient), therefore, almost (resp. hardly) any deviation from this criterion caused by point-mutations triggers recognition of the point-mutant by the T cell; this results in higher (resp. lower) point-mutant recognition probability at lower (resp. higher) T cell negative selection survival probability. 

Next, we compare the results at different $N_q$. This is a bit tricky, because fixing the negative selection probability leads to different thresholds $U_n$ at different training set sizes. This accounts for a large part but not all of the difference in the curves seen in Fig. \ref{fig:PMRP-Nq}; see SI Fig S8. By increasing the size of the negative-selecting repertoire $N_q$, a mature T cell's sense for self-antigen resemblance broadens; thus leading to higher tolerance (less detectability) for point-mutants at higher $N_q$ values. 

Another feature impacting point-mutant recognition probability that stems from incorporating contact maps into the model, pertains to the site in the pMHC sequence of the mutated AA. As can be seen in the contact maps in Fig. \ref{fig:CMs}, some pMHC AAs make more significant contacts with TCR AAs than other pMHC AAs. In the case of 3QIB's CDR3$\alpha$-pMHC contact map (top left of Fig. \ref{fig:CMs}), the number of non-vanishing contacts for a particular pMHC AA ranges from 1 (sparse-contact site) to 5 (high-contact site), with an averaged 3.06 TCR AAs in contact by the 7 pMHC AAs with non-vanishing contacts. Accordingly, a point-mutant $\tilde{q}$ with its mutation occurring in a sparse-contact site (resp. high-contact site) bears higher (resp. lower) resemblance with the non-mutant $q$ for a T cell. This effect clearly should impact the point-mutant recognition probability, with high-contact site point-mutants having higher recognition probability than their sparse-contact counterparts, and point-mutants with randomly chosen mutation sites having recognition probability somewhere in between the aforementioned two. We investigated this idea by running three simulations as explained in the paragraph above, but with the additional constraint that in each round of simulations the mutated site was: one, always a high-contact site; two, always a sparse-contact site; and three, randomly chosen. The negative-selection repertoire was fixed at $N_q = 10^4$. The point-mutant recognition probability of these simulations are shown in Fig. \ref{fig:PMRP-Contacts} and exhibit agreement with the expected behavior.

The aforementioned RICE framework cannot adequately distinguish high contact sites from sparse ones on either the TCR or pMHC amino acid sequences. RICE's prediction for neo-epitope recognition probability therefore represent fixed estimates for a typical `one-contact' mutation.  On the other hand, this new approach enables a quantitative estimate of this obvious dependence.  This aligns with previous strategies calling for mutations to target TCR-facing peptide amino acids; see for example \cite{Chowell2015}, \cite{Shang2009}  

\section{Conclusions}\label{sec:conclusions}

In this manuscript, we considered the role of a non-trivial contact map acting as a template for the explicit interactions between the TCR and pMHC AA sequences. This approach is a compromise between making an arbitrary rule as to how these sequences interact (for example, assuming only diagonal coupling as done in previous models) or using measured crystal structure for each considered pair, an obvious impossibility for anything resembling a large repertoire undergoing negative selection. The formulation isolates contributions from spatial conformation of CDR3 loops and pMHC complexes into these contact maps, while remaining features are encapsulated in energy coefficient matrices. Although all the analysis here was done using randomly generated energy matrices, serving as a baseline ``toy" model, the methodology is not restricted to such a choice, and other energy matrices such as the hydrophobicity-driven MJ matrix \cite{Miyazawa1985, Miyazawa1996}, or data-driven matrices \cite{Lin2021} can be used instead.

We observed that the inclusion of contact maps gave rise to several features impacting the variance of the TCR-pMHC binding energy: a density-related one, as the number of non-vanishing contacts correlates with increased variance; and a topology-related one, in which the repeat structure of the AAs in CDR3-loops' and in pMHC-complexes' sequences also skews the variance, with additional repeats correlating with increased variance. These changes in variance also affect negative selection recognition probabilities, with larger variances driving higher recognition probabilities.  The proposed generalization is therefore useful for characterizing the distributional behavior of TCR systems with a relatively fixed contact structure. Given that even at fixed MHC allele there are likely to be several distinct spatial conformations that can give rise to effective binding, a full treatment of the repertoire should include finding the set of templates that gives rise to the largest possible binding for the sequences under consideration. This extension will be reported on elsewhere.

Another influence of the topology of the contact map manifest in the recognition probability of point-mutated antigens by T cells that have been negatively selected. Here, some pMHC-AAs have higher number of non-vanishing contacts with TCR-AAs, that upon mutation make the antigen to be perceived more like foreign by the T cells than when mutating pMHC-AAs with fewer non-vanishing contacts; this results in higher recognition probability of high-contact site point-mutants. Conversely, this notion can provide at least some information about which mutations in a previously detected peptide could prevent the detection of an evolved virus by memory T cells generated in an earlier infection. Data to this effect is now becoming available in the context of COVID-19-specific T cells in never infected individuals resulting from prior responses to other endemic coronaviruses \cite{braun2020}. 

As seen here, the problem of dissecting the generation and functioning of the post-selection T cell repertoire is incredibly complex, even utilizing a number of vastly simplifying assumptions. The full problem requires attention to biases in the generation of the na\"ive repertoire \cite{murugan2012}, inclusion of a set of different MHC alleles for different individuals, a better handle on the statistical properties of the negative selection training set, and of course the full range of molecular biophysics effects that contribute to binding energy and on-off kinetics. These cannot all be included in any useful theoretical model. By isolating and improving our understanding of the effects of specific contact geometries, we hope to build intuition for how different aspects of this complex system contribute to different functional aspects of the full T cell arm of adaptive immunity.\\

\section{Acknowledgments}
The authors would like to thank Dr. Michael E. Birnbaum for fruitful discussion on systems-level TCR-antigen specificity. This work was supported by National Science Foundation (NSF) grant NSF PHY-2019745 (Center for Theoretical Biological Physics).

\bibliography{References-Immunologyv2} 

\providecommand{\noopsort}[1]{}\providecommand{\singleletter}[1]{#1}%
\begin{thebibliography}{36}%
\makeatletter
\providecommand \@ifxundefined [1]{%
 \@ifx{#1\undefined}
}%
\providecommand \@ifnum [1]{%
 \ifnum #1\expandafter \@firstoftwo
 \else \expandafter \@secondoftwo
 \fi
}%
\providecommand \@ifx [1]{%
 \ifx #1\expandafter \@firstoftwo
 \else \expandafter \@secondoftwo
 \fi
}%
\providecommand \natexlab [1]{#1}%
\providecommand \enquote  [1]{``#1''}%
\providecommand \bibnamefont  [1]{#1}%
\providecommand \bibfnamefont [1]{#1}%
\providecommand \citenamefont [1]{#1}%
\providecommand \href@noop [0]{\@secondoftwo}%
\providecommand \href [0]{\begingroup \@sanitize@url \@href}%
\providecommand \@href[1]{\@@startlink{#1}\@@href}%
\providecommand \@@href[1]{\endgroup#1\@@endlink}%
\providecommand \@sanitize@url [0]{\catcode `\\12\catcode `\$12\catcode
  `\&12\catcode `\#12\catcode `\^12\catcode `\_12\catcode `\%12\relax}%
\providecommand \@@startlink[1]{}%
\providecommand \@@endlink[0]{}%
\providecommand \url  [0]{\begingroup\@sanitize@url \@url }%
\providecommand \@url [1]{\endgroup\@href {#1}{\urlprefix }}%
\providecommand \urlprefix  [0]{URL }%
\providecommand \Eprint [0]{\href }%
\providecommand \doibase [0]{https://doi.org/}%
\providecommand \selectlanguage [0]{\@gobble}%
\providecommand \bibinfo  [0]{\@secondoftwo}%
\providecommand \bibfield  [0]{\@secondoftwo}%
\providecommand \translation [1]{[#1]}%
\providecommand \BibitemOpen [0]{}%
\providecommand \bibitemStop [0]{}%
\providecommand \bibitemNoStop [0]{.\EOS\space}%
\providecommand \EOS [0]{\spacefactor3000\relax}%
\providecommand \BibitemShut  [1]{\csname bibitem#1\endcsname}%
\let\auto@bib@innerbib\@empty
\bibitem [{\citenamefont {Ding}\ \emph {et~al.}(2012)\citenamefont {Ding},
  \citenamefont {Ley}, \citenamefont {Larson}, \citenamefont {Miller},
  \citenamefont {Koboldt}, \citenamefont {Welch}, \citenamefont {Ritchey},
  \citenamefont {Young}, \citenamefont {Lamprecht}, \citenamefont {McLellan}
  \emph {et~al.}}]{Ding2012}%
  \BibitemOpen
  \bibfield  {author} {\bibinfo {author} {\bibfnamefont {L.}~\bibnamefont
  {Ding}}, \bibinfo {author} {\bibfnamefont {T.~J.}\ \bibnamefont {Ley}},
  \bibinfo {author} {\bibfnamefont {D.~E.}\ \bibnamefont {Larson}}, \bibinfo
  {author} {\bibfnamefont {C.~A.}\ \bibnamefont {Miller}}, \bibinfo {author}
  {\bibfnamefont {D.~C.}\ \bibnamefont {Koboldt}}, \bibinfo {author}
  {\bibfnamefont {J.~S.}\ \bibnamefont {Welch}}, \bibinfo {author}
  {\bibfnamefont {J.~K.}\ \bibnamefont {Ritchey}}, \bibinfo {author}
  {\bibfnamefont {M.~A.}\ \bibnamefont {Young}}, \bibinfo {author}
  {\bibfnamefont {T.}~\bibnamefont {Lamprecht}}, \bibinfo {author}
  {\bibfnamefont {M.~D.}\ \bibnamefont {McLellan}}, \emph {et~al.},\ }\bibfield
   {title} {\bibinfo {title} {Clonal evolution in relapsed acute myeloid
  leukaemia revealed by whole-genome sequencing},\ }\href
  {https://doi.org/10.1038/nature10738} {\bibfield  {journal} {\bibinfo
  {journal} {Nature}\ }\textbf {\bibinfo {volume} {481}},\ \bibinfo {pages}
  {506} (\bibinfo {year} {2012})}\BibitemShut {NoStop}%
\bibitem [{\citenamefont {Robinson}\ \emph {et~al.}(2016)\citenamefont
  {Robinson}, \citenamefont {Soormally}, \citenamefont {Hayhurst},\ and\
  \citenamefont {Marsh}}]{Robinson2016}%
  \BibitemOpen
  \bibfield  {author} {\bibinfo {author} {\bibfnamefont {J.}~\bibnamefont
  {Robinson}}, \bibinfo {author} {\bibfnamefont {A.~R.}\ \bibnamefont
  {Soormally}}, \bibinfo {author} {\bibfnamefont {J.~D.}\ \bibnamefont
  {Hayhurst}},\ and\ \bibinfo {author} {\bibfnamefont {S.~G.~E.}\ \bibnamefont
  {Marsh}},\ }\bibfield  {title} {\bibinfo {title} {The ipd-imgt/hla database -
  new developments in reporting hla variation},\ }\href
  {https://doi.org/10.1016/j.humimm.2016.01.020} {\bibfield  {journal}
  {\bibinfo  {journal} {Human immunology}\ }\textbf {\bibinfo {volume} {77}},\
  \bibinfo {pages} {233} (\bibinfo {year} {2016})}\BibitemShut {NoStop}%
\bibitem [{\citenamefont {Schumacher}\ and\ \citenamefont
  {Schreiber}(2015)}]{Schumacher2015}%
  \BibitemOpen
  \bibfield  {author} {\bibinfo {author} {\bibfnamefont {T.~N.}\ \bibnamefont
  {Schumacher}}\ and\ \bibinfo {author} {\bibfnamefont {R.~D.}\ \bibnamefont
  {Schreiber}},\ }\bibfield  {title} {\bibinfo {title} {Neoantigens in cancer
  immunotherapy},\ }\href {https://doi.org/10.1126/science.aaa4971} {\bibfield
  {journal} {\bibinfo  {journal} {Science}\ }\textbf {\bibinfo {volume}
  {348}},\ \bibinfo {pages} {69} (\bibinfo {year} {2015})}\BibitemShut
  {NoStop}%
\bibitem [{\citenamefont {Verdegaal}\ \emph {et~al.}(2016)\citenamefont
  {Verdegaal}, \citenamefont {de~Miranda}, \citenamefont {Visser},
  \citenamefont {Harryvan}, \citenamefont {van Buuren}, \citenamefont
  {Andersen}, \citenamefont {Hadrup}, \citenamefont {van~der Minne},
  \citenamefont {Schotte}, \citenamefont {Spits} \emph
  {et~al.}}]{Verdegaal2016}%
  \BibitemOpen
  \bibfield  {author} {\bibinfo {author} {\bibfnamefont {E.~M.~E.}\
  \bibnamefont {Verdegaal}}, \bibinfo {author} {\bibfnamefont {N.~F. C.~C.}\
  \bibnamefont {de~Miranda}}, \bibinfo {author} {\bibfnamefont
  {M.}~\bibnamefont {Visser}}, \bibinfo {author} {\bibfnamefont
  {T.}~\bibnamefont {Harryvan}}, \bibinfo {author} {\bibfnamefont {M.~M.}\
  \bibnamefont {van Buuren}}, \bibinfo {author} {\bibfnamefont {R.~S.}\
  \bibnamefont {Andersen}}, \bibinfo {author} {\bibfnamefont {S.~R.}\
  \bibnamefont {Hadrup}}, \bibinfo {author} {\bibfnamefont {C.~E.}\
  \bibnamefont {van~der Minne}}, \bibinfo {author} {\bibfnamefont
  {R.}~\bibnamefont {Schotte}}, \bibinfo {author} {\bibfnamefont
  {H.}~\bibnamefont {Spits}}, \emph {et~al.},\ }\bibfield  {title} {\bibinfo
  {title} {Neoantigen landscape dynamics during human melanoma-t cell
  interactions},\ }\href {https://doi.org/10.1038/nature18945} {\bibfield
  {journal} {\bibinfo  {journal} {Nature}\ }\textbf {\bibinfo {volume} {536}},\
  \bibinfo {pages} {91} (\bibinfo {year} {2016})}\BibitemShut {NoStop}%
\bibitem [{\citenamefont {Das}\ \emph {et~al.}(2015)\citenamefont {Das},
  \citenamefont {Feng}, \citenamefont {Mallis}, \citenamefont {Li},
  \citenamefont {Keskin}, \citenamefont {Hussey}, \citenamefont {Brady},
  \citenamefont {Wang}, \citenamefont {Wagner}, \citenamefont {Reinherz} \emph
  {et~al.}}]{Das2015}%
  \BibitemOpen
  \bibfield  {author} {\bibinfo {author} {\bibfnamefont {D.~K.}\ \bibnamefont
  {Das}}, \bibinfo {author} {\bibfnamefont {Y.}~\bibnamefont {Feng}}, \bibinfo
  {author} {\bibfnamefont {R.~J.}\ \bibnamefont {Mallis}}, \bibinfo {author}
  {\bibfnamefont {X.}~\bibnamefont {Li}}, \bibinfo {author} {\bibfnamefont
  {D.~B.}\ \bibnamefont {Keskin}}, \bibinfo {author} {\bibfnamefont {R.~E.}\
  \bibnamefont {Hussey}}, \bibinfo {author} {\bibfnamefont {S.~K.}\
  \bibnamefont {Brady}}, \bibinfo {author} {\bibfnamefont {J.-H.}\ \bibnamefont
  {Wang}}, \bibinfo {author} {\bibfnamefont {G.}~\bibnamefont {Wagner}},
  \bibinfo {author} {\bibfnamefont {E.~L.}\ \bibnamefont {Reinherz}}, \emph
  {et~al.},\ }\bibfield  {title} {\bibinfo {title} {Force-dependent transition
  in the t-cell receptor $\beta$-subunit allosterically regulates peptide
  discrimination and pmhc bond lifetime},\ }\href@noop {} {\bibfield  {journal}
  {\bibinfo  {journal} {Proceedings of the National Academy of Sciences}\
  }\textbf {\bibinfo {volume} {112}},\ \bibinfo {pages} {1517} (\bibinfo {year}
  {2015})}\BibitemShut {NoStop}%
\bibitem [{\citenamefont {Fran{\c{c}}ois}\ and\ \citenamefont
  {Altan-Bonnet}(2016)}]{Francois2016}%
  \BibitemOpen
  \bibfield  {author} {\bibinfo {author} {\bibfnamefont {P.}~\bibnamefont
  {Fran{\c{c}}ois}}\ and\ \bibinfo {author} {\bibfnamefont {G.}~\bibnamefont
  {Altan-Bonnet}},\ }\bibfield  {title} {\bibinfo {title} {The case for
  absolute ligand discrimination: modeling information processing and decision
  by immune t cells},\ }\href@noop {} {\bibfield  {journal} {\bibinfo
  {journal} {Journal of Statistical Physics}\ }\textbf {\bibinfo {volume}
  {162}},\ \bibinfo {pages} {1130} (\bibinfo {year} {2016})}\BibitemShut
  {NoStop}%
\bibitem [{\citenamefont {Alam}\ \emph {et~al.}(1996)\citenamefont {Alam},
  \citenamefont {Travers}, \citenamefont {Wung}, \citenamefont {Nasholds},
  \citenamefont {Redpath}, \citenamefont {Jameson},\ and\ \citenamefont
  {Gascoigne}}]{Alam1996}%
  \BibitemOpen
  \bibfield  {author} {\bibinfo {author} {\bibfnamefont {S.~M.}\ \bibnamefont
  {Alam}}, \bibinfo {author} {\bibfnamefont {P.~J.}\ \bibnamefont {Travers}},
  \bibinfo {author} {\bibfnamefont {J.~L.}\ \bibnamefont {Wung}}, \bibinfo
  {author} {\bibfnamefont {W.}~\bibnamefont {Nasholds}}, \bibinfo {author}
  {\bibfnamefont {S.}~\bibnamefont {Redpath}}, \bibinfo {author} {\bibfnamefont
  {S.~C.}\ \bibnamefont {Jameson}},\ and\ \bibinfo {author} {\bibfnamefont
  {N.~R.~J.}\ \bibnamefont {Gascoigne}},\ }\bibfield  {title} {\bibinfo {title}
  {T-cell-receptor affinity and thymocyte positive selection},\ }\href
  {https://doi.org/10.1038/381616a0} {\bibfield  {journal} {\bibinfo  {journal}
  {Nature}\ }\textbf {\bibinfo {volume} {381}},\ \bibinfo {pages} {616}
  (\bibinfo {year} {1996})}\BibitemShut {NoStop}%
\bibitem [{\citenamefont {Krogsgaard}\ and\ \citenamefont
  {Davis}(2005)}]{Krogsgaard2005}%
  \BibitemOpen
  \bibfield  {author} {\bibinfo {author} {\bibfnamefont {M.}~\bibnamefont
  {Krogsgaard}}\ and\ \bibinfo {author} {\bibfnamefont {M.~M.}\ \bibnamefont
  {Davis}},\ }\bibfield  {title} {\bibinfo {title} {How t cells 'see'
  antigen},\ }\href {https://doi.org/10.1038/ni117} {\bibfield  {journal}
  {\bibinfo  {journal} {Nature immunology}\ }\textbf {\bibinfo {volume} {6}},\
  \bibinfo {pages} {239} (\bibinfo {year} {2005})}\BibitemShut {NoStop}%
\bibitem [{\citenamefont {Arstila}\ \emph {et~al.}(1999)\citenamefont
  {Arstila}, \citenamefont {Casrouge}, \citenamefont {Baron}, \citenamefont
  {Even}, \citenamefont {Kanellopoulos},\ and\ \citenamefont
  {Kourilsky}}]{Arstila1999}%
  \BibitemOpen
  \bibfield  {author} {\bibinfo {author} {\bibfnamefont {T.~P.}\ \bibnamefont
  {Arstila}}, \bibinfo {author} {\bibfnamefont {A.}~\bibnamefont {Casrouge}},
  \bibinfo {author} {\bibfnamefont {V.}~\bibnamefont {Baron}}, \bibinfo
  {author} {\bibfnamefont {J.}~\bibnamefont {Even}}, \bibinfo {author}
  {\bibfnamefont {J.}~\bibnamefont {Kanellopoulos}},\ and\ \bibinfo {author}
  {\bibfnamefont {P.}~\bibnamefont {Kourilsky}},\ }\bibfield  {title} {\bibinfo
  {title} {A direct estimate of the human $\alpha$ $\beta$ t cell receptor
  diversity},\ }\href@noop {} {\bibfield  {journal} {\bibinfo  {journal}
  {Science}\ }\textbf {\bibinfo {volume} {286}},\ \bibinfo {pages} {958}
  (\bibinfo {year} {1999})}\BibitemShut {NoStop}%
\bibitem [{\citenamefont {De~Boer}\ and\ \citenamefont
  {Perelson}(1993)}]{DeBoer1993}%
  \BibitemOpen
  \bibfield  {author} {\bibinfo {author} {\bibfnamefont {R.~J.}\ \bibnamefont
  {De~Boer}}\ and\ \bibinfo {author} {\bibfnamefont {A.~S.}\ \bibnamefont
  {Perelson}},\ }\bibfield  {title} {\bibinfo {title} {How diverse should the
  immune system be?},\ }\href {https://doi.org/10.1098/rspb.1993.0062}
  {\bibfield  {journal} {\bibinfo  {journal} {Proceedings. Biological
  Sciences}\ }\textbf {\bibinfo {volume} {252}},\ \bibinfo {pages} {171}
  (\bibinfo {year} {1993})}\BibitemShut {NoStop}%
\bibitem [{\citenamefont {Yates}(2014)}]{Yates2014}%
  \BibitemOpen
  \bibfield  {author} {\bibinfo {author} {\bibfnamefont {A.}~\bibnamefont
  {Yates}},\ }\bibfield  {title} {\bibinfo {title} {Theories and quantification
  of thymic selection},\ }\href {https://doi.org/10.3389/fimmu.2014.00013}
  {\bibfield  {journal} {\bibinfo  {journal} {Frontiers in Immunology}\
  }\textbf {\bibinfo {volume} {5}},\ \bibinfo {pages} {13} (\bibinfo {year}
  {2014})}\BibitemShut {NoStop}%
\bibitem [{\citenamefont {Detours}\ and\ \citenamefont
  {Perelson}(1999)}]{Detours1999}%
  \BibitemOpen
  \bibfield  {author} {\bibinfo {author} {\bibfnamefont {V.}~\bibnamefont
  {Detours}}\ and\ \bibinfo {author} {\bibfnamefont {A.~S.}\ \bibnamefont
  {Perelson}},\ }\bibfield  {title} {\bibinfo {title} {Explaining high
  alloreactivity as a quantitative consequence of affinity-driven thymocyte
  selection},\ }\href@noop {} {\bibfield  {journal} {\bibinfo  {journal}
  {Proceedings of the National Academy of Sciences}\ }\textbf {\bibinfo
  {volume} {96}},\ \bibinfo {pages} {5153} (\bibinfo {year}
  {1999})}\BibitemShut {NoStop}%
\bibitem [{\citenamefont {Klein}\ \emph {et~al.}(2014)\citenamefont {Klein},
  \citenamefont {Kyewski}, \citenamefont {Allen},\ and\ \citenamefont
  {Hogquist}}]{Klein2014}%
  \BibitemOpen
  \bibfield  {author} {\bibinfo {author} {\bibfnamefont {L.}~\bibnamefont
  {Klein}}, \bibinfo {author} {\bibfnamefont {B.}~\bibnamefont {Kyewski}},
  \bibinfo {author} {\bibfnamefont {P.~M.}\ \bibnamefont {Allen}},\ and\
  \bibinfo {author} {\bibfnamefont {K.~A.}\ \bibnamefont {Hogquist}},\
  }\bibfield  {title} {\bibinfo {title} {Positive and negative selection of the
  t cell repertoire: what thymocytes see (and don't see)},\ }\href
  {https://doi.org/10.1038/nri3667} {\bibfield  {journal} {\bibinfo  {journal}
  {Nature reviews. Immunology}\ }\textbf {\bibinfo {volume} {14}},\ \bibinfo
  {pages} {377} (\bibinfo {year} {2014})}\BibitemShut {NoStop}%
\bibitem [{\citenamefont {Lanzarotti}\ \emph {et~al.}(2018)\citenamefont
  {Lanzarotti}, \citenamefont {Marcatili},\ and\ \citenamefont
  {Nielsen}}]{Lanzarotti2011}%
  \BibitemOpen
  \bibfield  {author} {\bibinfo {author} {\bibfnamefont {E.}~\bibnamefont
  {Lanzarotti}}, \bibinfo {author} {\bibfnamefont {P.}~\bibnamefont
  {Marcatili}},\ and\ \bibinfo {author} {\bibfnamefont {M.}~\bibnamefont
  {Nielsen}},\ }\bibfield  {title} {\bibinfo {title} {Identification of the
  cognate peptide-mhc target of t cell receptors using molecular modeling and
  force field scoring},\ }\href
  {https://doi.org/https://doi.org/10.1016/j.molimm.2017.12.019} {\bibfield
  {journal} {\bibinfo  {journal} {Molecular Immunology}\ }\textbf {\bibinfo
  {volume} {94}},\ \bibinfo {pages} {91 } (\bibinfo {year} {2018})}\BibitemShut
  {NoStop}%
\bibitem [{\citenamefont {Newell}\ \emph {et~al.}(2011)\citenamefont {Newell},
  \citenamefont {Ely}, \citenamefont {Kruse}, \citenamefont {Reay},
  \citenamefont {Rodriguez}, \citenamefont {Lin}, \citenamefont {Kuhns},
  \citenamefont {Garcia},\ and\ \citenamefont {Davis}}]{Newell2011}%
  \BibitemOpen
  \bibfield  {author} {\bibinfo {author} {\bibfnamefont {E.~W.}\ \bibnamefont
  {Newell}}, \bibinfo {author} {\bibfnamefont {L.~K.}\ \bibnamefont {Ely}},
  \bibinfo {author} {\bibfnamefont {A.~C.}\ \bibnamefont {Kruse}}, \bibinfo
  {author} {\bibfnamefont {P.~A.}\ \bibnamefont {Reay}}, \bibinfo {author}
  {\bibfnamefont {S.~N.}\ \bibnamefont {Rodriguez}}, \bibinfo {author}
  {\bibfnamefont {A.~E.}\ \bibnamefont {Lin}}, \bibinfo {author} {\bibfnamefont
  {M.~S.}\ \bibnamefont {Kuhns}}, \bibinfo {author} {\bibfnamefont {K.~C.}\
  \bibnamefont {Garcia}},\ and\ \bibinfo {author} {\bibfnamefont {M.~M.}\
  \bibnamefont {Davis}},\ }\bibfield  {title} {\bibinfo {title} {Structural
  basis of specificity and cross-reactivity in t cell receptors specific for
  cytochrome c{\textendash}i-ek},\ }\href
  {https://doi.org/10.4049/jimmunol.1100197} {\bibfield  {journal} {\bibinfo
  {journal} {The Journal of Immunology}\ }\textbf {\bibinfo {volume} {186}},\
  \bibinfo {pages} {5823} (\bibinfo {year} {2011})}\BibitemShut {NoStop}%
\bibitem [{\citenamefont {Baker}\ \emph {et~al.}(2012)\citenamefont {Baker},
  \citenamefont {Scott}, \citenamefont {Blevins},\ and\ \citenamefont
  {Hawse}}]{Baker2012}%
  \BibitemOpen
  \bibfield  {author} {\bibinfo {author} {\bibfnamefont {B.~M.}\ \bibnamefont
  {Baker}}, \bibinfo {author} {\bibfnamefont {D.~R.}\ \bibnamefont {Scott}},
  \bibinfo {author} {\bibfnamefont {S.~J.}\ \bibnamefont {Blevins}},\ and\
  \bibinfo {author} {\bibfnamefont {W.~F.}\ \bibnamefont {Hawse}},\ }\bibfield
  {title} {\bibinfo {title} {Structural and dynamic control of t-cell receptor
  specificity, cross-reactivity, and binding mechanism},\ }\href
  {https://doi.org/https://doi.org/10.1111/j.1600-065X.2012.01165.x} {\bibfield
   {journal} {\bibinfo  {journal} {Immunological Reviews}\ }\textbf {\bibinfo
  {volume} {250}},\ \bibinfo {pages} {10} (\bibinfo {year} {2012})}\BibitemShut
  {NoStop}%
\bibitem [{\citenamefont {Colf}\ \emph {et~al.}(2007)\citenamefont {Colf},
  \citenamefont {Bankovich}, \citenamefont {Hanick}, \citenamefont {Bowerman},
  \citenamefont {Jones}, \citenamefont {Kranz},\ and\ \citenamefont
  {Garcia}}]{Colf2007}%
  \BibitemOpen
  \bibfield  {author} {\bibinfo {author} {\bibfnamefont {L.~A.}\ \bibnamefont
  {Colf}}, \bibinfo {author} {\bibfnamefont {A.~J.}\ \bibnamefont {Bankovich}},
  \bibinfo {author} {\bibfnamefont {N.~A.}\ \bibnamefont {Hanick}}, \bibinfo
  {author} {\bibfnamefont {N.~A.}\ \bibnamefont {Bowerman}}, \bibinfo {author}
  {\bibfnamefont {L.~L.}\ \bibnamefont {Jones}}, \bibinfo {author}
  {\bibfnamefont {D.}~\bibnamefont {Kranz}},\ and\ \bibinfo {author}
  {\bibfnamefont {K.~C.}\ \bibnamefont {Garcia}},\ }\bibfield  {title}
  {\bibinfo {title} {How a single t cell receptor recognizes both self and
  foreign mhc},\ }\href
  {https://doi.org/https://doi.org/10.1016/j.cell.2007.01.048} {\bibfield
  {journal} {\bibinfo  {journal} {Cell}\ }\textbf {\bibinfo {volume} {129}},\
  \bibinfo {pages} {135 } (\bibinfo {year} {2007})}\BibitemShut {NoStop}%
\bibitem [{\citenamefont {Ko{\v s}mrlj}\ \emph {et~al.}(2008)\citenamefont
  {Ko{\v s}mrlj}, \citenamefont {Jha}, \citenamefont {Huseby}, \citenamefont
  {Kardar},\ and\ \citenamefont {Chakraborty}}]{Kosmrlj2008}%
  \BibitemOpen
  \bibfield  {author} {\bibinfo {author} {\bibfnamefont {A.}~\bibnamefont
  {Ko{\v s}mrlj}}, \bibinfo {author} {\bibfnamefont {A.~K.}\ \bibnamefont
  {Jha}}, \bibinfo {author} {\bibfnamefont {E.~S.}\ \bibnamefont {Huseby}},
  \bibinfo {author} {\bibfnamefont {M.}~\bibnamefont {Kardar}},\ and\ \bibinfo
  {author} {\bibfnamefont {A.~K.}\ \bibnamefont {Chakraborty}},\ }\bibfield
  {title} {\bibinfo {title} {How the thymus designs antigen-specific and
  self-tolerant t cell receptor sequences},\ }\href
  {https://doi.org/10.1073/pnas.0808081105} {\bibfield  {journal} {\bibinfo
  {journal} {Proceedings of the National Academy of Sciences}\ }\textbf
  {\bibinfo {volume} {105}},\ \bibinfo {pages} {16671} (\bibinfo {year}
  {2008})}\BibitemShut {NoStop}%
\bibitem [{\citenamefont {Ko\ifmmode~\check{s}\else \v{s}\fi{}mrlj}\ \emph
  {et~al.}(2009)\citenamefont {Ko\ifmmode~\check{s}\else \v{s}\fi{}mrlj},
  \citenamefont {Chakraborty}, \citenamefont {Kardar},\ and\ \citenamefont
  {Shakhnovich}}]{Kosmrlj2009}%
  \BibitemOpen
  \bibfield  {author} {\bibinfo {author} {\bibfnamefont {A.}~\bibnamefont
  {Ko\ifmmode~\check{s}\else \v{s}\fi{}mrlj}}, \bibinfo {author} {\bibfnamefont
  {A.~K.}\ \bibnamefont {Chakraborty}}, \bibinfo {author} {\bibfnamefont
  {M.}~\bibnamefont {Kardar}},\ and\ \bibinfo {author} {\bibfnamefont {E.~I.}\
  \bibnamefont {Shakhnovich}},\ }\bibfield  {title} {\bibinfo {title} {Thymic
  selection of t-cell receptors as an extreme value problem},\ }\href
  {https://doi.org/10.1103/PhysRevLett.103.068103} {\bibfield  {journal}
  {\bibinfo  {journal} {Phys. Rev. Lett.}\ }\textbf {\bibinfo {volume} {103}},\
  \bibinfo {pages} {068103} (\bibinfo {year} {2009})}\BibitemShut {NoStop}%
\bibitem [{\citenamefont {Chakraborty}\ and\ \citenamefont
  {Košmrlj}(2010)}]{Chakraborty2010}%
  \BibitemOpen
  \bibfield  {author} {\bibinfo {author} {\bibfnamefont {A.~K.}\ \bibnamefont
  {Chakraborty}}\ and\ \bibinfo {author} {\bibfnamefont {A.}~\bibnamefont
  {Košmrlj}},\ }\bibfield  {title} {\bibinfo {title} {Statistical mechanical
  concepts in immunology},\ }\href
  {https://doi.org/10.1146/annurev.physchem.59.032607.093537} {\bibfield
  {journal} {\bibinfo  {journal} {Annual Review of Physical Chemistry}\
  }\textbf {\bibinfo {volume} {61}},\ \bibinfo {pages} {283} (\bibinfo {year}
  {2010})},\ \bibinfo {note} {pMID: 20367082}\BibitemShut {NoStop}%
\bibitem [{\citenamefont {George}\ \emph {et~al.}(2017)\citenamefont {George},
  \citenamefont {Kessler},\ and\ \citenamefont {Levine}}]{George2017}%
  \BibitemOpen
  \bibfield  {author} {\bibinfo {author} {\bibfnamefont {J.~T.}\ \bibnamefont
  {George}}, \bibinfo {author} {\bibfnamefont {D.~A.}\ \bibnamefont
  {Kessler}},\ and\ \bibinfo {author} {\bibfnamefont {H.}~\bibnamefont
  {Levine}},\ }\bibfield  {title} {\bibinfo {title} {Effects of thymic
  selection on t cell recognition of foreign and tumor antigenic peptides},\
  }\href {https://doi.org/10.1073/pnas.1708573114} {\bibfield  {journal}
  {\bibinfo  {journal} {Proceedings of the National Academy of Sciences}\
  }\textbf {\bibinfo {volume} {114}},\ \bibinfo {pages} {E7875} (\bibinfo
  {year} {2017})}\BibitemShut {NoStop}%
\bibitem [{\citenamefont {Wortel}\ \emph {et~al.}(2020)\citenamefont {Wortel},
  \citenamefont {Ke{\c{s}}mir}, \citenamefont {de~Boer}, \citenamefont
  {Mandl},\ and\ \citenamefont {Textor}}]{Wortel2020}%
  \BibitemOpen
  \bibfield  {author} {\bibinfo {author} {\bibfnamefont {I.}~\bibnamefont
  {Wortel}}, \bibinfo {author} {\bibfnamefont {C.}~\bibnamefont
  {Ke{\c{s}}mir}}, \bibinfo {author} {\bibfnamefont {R.~J.}\ \bibnamefont
  {de~Boer}}, \bibinfo {author} {\bibfnamefont {J.~N.}\ \bibnamefont {Mandl}},\
  and\ \bibinfo {author} {\bibfnamefont {J.}~\bibnamefont {Textor}},\
  }\bibfield  {title} {\bibinfo {title} {Is t cell negative selection a
  learning algorithm?},\ }\href@noop {} {\bibfield  {journal} {\bibinfo
  {journal} {Cells}\ }\textbf {\bibinfo {volume} {9}},\ \bibinfo {pages} {690}
  (\bibinfo {year} {2020})}\BibitemShut {NoStop}%
\bibitem [{\citenamefont {Ko{\v s}mrlj}\ \emph {et~al.}(2010)\citenamefont
  {Ko{\v s}mrlj}, \citenamefont {Read}, \citenamefont {Qi}, \citenamefont
  {Allen}, \citenamefont {Altfeld}, \citenamefont {Deeks}, \citenamefont
  {Pereyra}, \citenamefont {Carrington}, \citenamefont {Walker},\ and\
  \citenamefont {Chakraborty}}]{Kosmrlj2010}%
  \BibitemOpen
  \bibfield  {author} {\bibinfo {author} {\bibfnamefont {A.}~\bibnamefont
  {Ko{\v s}mrlj}}, \bibinfo {author} {\bibfnamefont {E.~L.}\ \bibnamefont
  {Read}}, \bibinfo {author} {\bibfnamefont {Y.}~\bibnamefont {Qi}}, \bibinfo
  {author} {\bibfnamefont {T.~M.}\ \bibnamefont {Allen}}, \bibinfo {author}
  {\bibfnamefont {M.}~\bibnamefont {Altfeld}}, \bibinfo {author} {\bibfnamefont
  {S.~G.}\ \bibnamefont {Deeks}}, \bibinfo {author} {\bibfnamefont
  {F.}~\bibnamefont {Pereyra}}, \bibinfo {author} {\bibfnamefont
  {M.}~\bibnamefont {Carrington}}, \bibinfo {author} {\bibfnamefont {B.~D.}\
  \bibnamefont {Walker}},\ and\ \bibinfo {author} {\bibfnamefont {A.~K.}\
  \bibnamefont {Chakraborty}},\ }\bibfield  {title} {\bibinfo {title} {Effects
  of thymic selection of the t-cell repertoire on hla class i-associated
  control of hiv infection},\ }\href {https://doi.org/10.1038/nature08997}
  {\bibfield  {journal} {\bibinfo  {journal} {Nature}\ }\textbf {\bibinfo
  {volume} {465}},\ \bibinfo {pages} {350} (\bibinfo {year}
  {2010})}\BibitemShut {NoStop}%
\bibitem [{\citenamefont {Chen}\ \emph {et~al.}(2018)\citenamefont {Chen},
  \citenamefont {Chakraborty},\ and\ \citenamefont {Kardar}}]{Chen2018}%
  \BibitemOpen
  \bibfield  {author} {\bibinfo {author} {\bibfnamefont {H.}~\bibnamefont
  {Chen}}, \bibinfo {author} {\bibfnamefont {A.~K.}\ \bibnamefont
  {Chakraborty}},\ and\ \bibinfo {author} {\bibfnamefont {M.}~\bibnamefont
  {Kardar}},\ }\bibfield  {title} {\bibinfo {title} {How nonuniform contact
  profiles of t cell receptors modulate thymic selection outcomes},\ }\href
  {https://doi.org/10.1103/PhysRevE.97.032413} {\bibfield  {journal} {\bibinfo
  {journal} {Phys. Rev. E}\ }\textbf {\bibinfo {volume} {97}},\ \bibinfo
  {pages} {032413} (\bibinfo {year} {2018})}\BibitemShut {NoStop}%
\bibitem [{\citenamefont {Cole}\ \emph {et~al.}(2016)\citenamefont {Cole},
  \citenamefont {Bulek}, \citenamefont {Dolton}, \citenamefont {Schauenberg},
  \citenamefont {Szomolay}, \citenamefont {Rittase}, \citenamefont {Trimby},
  \citenamefont {Jothikumar}, \citenamefont {Fuller}, \citenamefont {Skowera}
  \emph {et~al.}}]{Cole2016}%
  \BibitemOpen
  \bibfield  {author} {\bibinfo {author} {\bibfnamefont {D.~K.}\ \bibnamefont
  {Cole}}, \bibinfo {author} {\bibfnamefont {A.~M.}\ \bibnamefont {Bulek}},
  \bibinfo {author} {\bibfnamefont {G.}~\bibnamefont {Dolton}}, \bibinfo
  {author} {\bibfnamefont {A.~J.}\ \bibnamefont {Schauenberg}}, \bibinfo
  {author} {\bibfnamefont {B.}~\bibnamefont {Szomolay}}, \bibinfo {author}
  {\bibfnamefont {W.}~\bibnamefont {Rittase}}, \bibinfo {author} {\bibfnamefont
  {A.}~\bibnamefont {Trimby}}, \bibinfo {author} {\bibfnamefont
  {P.}~\bibnamefont {Jothikumar}}, \bibinfo {author} {\bibfnamefont
  {A.}~\bibnamefont {Fuller}}, \bibinfo {author} {\bibfnamefont
  {A.}~\bibnamefont {Skowera}}, \emph {et~al.},\ }\bibfield  {title} {\bibinfo
  {title} {Hotspot autoimmune t cell receptor binding underlies pathogen and
  insulin peptide cross-reactivity},\ }\href {https://doi.org/10.1172/JCI85679}
  {\bibfield  {journal} {\bibinfo  {journal} {The Journal of Clinical
  Investigation}\ }\textbf {\bibinfo {volume} {126}},\ \bibinfo {pages} {2191}
  (\bibinfo {year} {2016})}\BibitemShut {NoStop}%
\bibitem [{\citenamefont {Sethi}\ \emph {et~al.}(2013)\citenamefont {Sethi},
  \citenamefont {Gordo}, \citenamefont {Schubert},\ and\ \citenamefont
  {Wucherpfennig}}]{Sethi2013}%
  \BibitemOpen
  \bibfield  {author} {\bibinfo {author} {\bibfnamefont {D.~K.}\ \bibnamefont
  {Sethi}}, \bibinfo {author} {\bibfnamefont {S.}~\bibnamefont {Gordo}},
  \bibinfo {author} {\bibfnamefont {D.~A.}\ \bibnamefont {Schubert}},\ and\
  \bibinfo {author} {\bibfnamefont {K.~W.}\ \bibnamefont {Wucherpfennig}},\
  }\bibfield  {title} {\bibinfo {title} {Crossreactivity of a human autoimmune
  tcr is dominated by a single tcr loop},\ }\href
  {https://doi.org/10.1038/ncomms3623} {\bibfield  {journal} {\bibinfo
  {journal} {Nature Communications}\ }\textbf {\bibinfo {volume} {4}},\
  \bibinfo {pages} {2623} (\bibinfo {year} {2013})}\BibitemShut {NoStop}%
\bibitem [{\citenamefont {Ting}\ \emph {et~al.}(2020)\citenamefont {Ting},
  \citenamefont {Dahal-Koirala}, \citenamefont {Kim}, \citenamefont {Qiao},
  \citenamefont {Neumann}, \citenamefont {Lundin}, \citenamefont {Petersen},
  \citenamefont {Reid}, \citenamefont {Sollid},\ and\ \citenamefont
  {Rossjohn}}]{Ting2020}%
  \BibitemOpen
  \bibfield  {author} {\bibinfo {author} {\bibfnamefont {Y.~T.}\ \bibnamefont
  {Ting}}, \bibinfo {author} {\bibfnamefont {S.}~\bibnamefont {Dahal-Koirala}},
  \bibinfo {author} {\bibfnamefont {H.~S.~K.}\ \bibnamefont {Kim}}, \bibinfo
  {author} {\bibfnamefont {S.-W.}\ \bibnamefont {Qiao}}, \bibinfo {author}
  {\bibfnamefont {R.~S.}\ \bibnamefont {Neumann}}, \bibinfo {author}
  {\bibfnamefont {K.~E.~A.}\ \bibnamefont {Lundin}}, \bibinfo {author}
  {\bibfnamefont {J.}~\bibnamefont {Petersen}}, \bibinfo {author}
  {\bibfnamefont {H.~H.}\ \bibnamefont {Reid}}, \bibinfo {author}
  {\bibfnamefont {L.~M.}\ \bibnamefont {Sollid}},\ and\ \bibinfo {author}
  {\bibfnamefont {J.}~\bibnamefont {Rossjohn}},\ }\bibfield  {title} {\bibinfo
  {title} {A molecular basis for the t cell response in hla-dq2.2 mediated
  celiac disease},\ }\href {https://doi.org/10.1073/pnas.1914308117} {\bibfield
   {journal} {\bibinfo  {journal} {Proceedings of the National Academy of
  Sciences}\ }\textbf {\bibinfo {volume} {117}},\ \bibinfo {pages} {3063}
  (\bibinfo {year} {2020})}\BibitemShut {NoStop}%
\bibitem [{\citenamefont {Lin}\ \emph {et~al.}(2021)\citenamefont {Lin},
  \citenamefont {George}, \citenamefont {Schafer}, \citenamefont {Ng~Chau},
  \citenamefont {Birnbaum}, \citenamefont {Clementi}, \citenamefont {Onuchic},\
  and\ \citenamefont {Levine}}]{Lin2021}%
  \BibitemOpen
  \bibfield  {author} {\bibinfo {author} {\bibfnamefont {X.}~\bibnamefont
  {Lin}}, \bibinfo {author} {\bibfnamefont {J.~T.}\ \bibnamefont {George}},
  \bibinfo {author} {\bibfnamefont {N.~P.}\ \bibnamefont {Schafer}}, \bibinfo
  {author} {\bibfnamefont {K.}~\bibnamefont {Ng~Chau}}, \bibinfo {author}
  {\bibfnamefont {M.~E.}\ \bibnamefont {Birnbaum}}, \bibinfo {author}
  {\bibfnamefont {C.}~\bibnamefont {Clementi}}, \bibinfo {author}
  {\bibfnamefont {J.~N.}\ \bibnamefont {Onuchic}},\ and\ \bibinfo {author}
  {\bibfnamefont {H.}~\bibnamefont {Levine}},\ }\bibfield  {title} {\bibinfo
  {title} {Rapid assessment of t-cell receptor specificity of the immune
  repertoire},\ }\href@noop {} {\bibfield  {journal} {\bibinfo  {journal}
  {Nature Computational Science}\ }\textbf {\bibinfo {volume} {1}},\ \bibinfo
  {pages} {362} (\bibinfo {year} {2021})}\BibitemShut {NoStop}%
\bibitem [{\citenamefont {Davtyan}\ \emph {et~al.}(2012)\citenamefont
  {Davtyan}, \citenamefont {Schafer}, \citenamefont {Zheng}, \citenamefont
  {Clementi}, \citenamefont {Wolynes},\ and\ \citenamefont
  {Papoian}}]{Davtyan2012}%
  \BibitemOpen
  \bibfield  {author} {\bibinfo {author} {\bibfnamefont {A.}~\bibnamefont
  {Davtyan}}, \bibinfo {author} {\bibfnamefont {N.~P.}\ \bibnamefont
  {Schafer}}, \bibinfo {author} {\bibfnamefont {W.}~\bibnamefont {Zheng}},
  \bibinfo {author} {\bibfnamefont {C.}~\bibnamefont {Clementi}}, \bibinfo
  {author} {\bibfnamefont {P.~G.}\ \bibnamefont {Wolynes}},\ and\ \bibinfo
  {author} {\bibfnamefont {G.~A.}\ \bibnamefont {Papoian}},\ }\bibfield
  {title} {\bibinfo {title} {Awsem-md: Protein structure prediction using
  coarse-grained physical potentials and bioinformatically based local
  structure biasing},\ }\href {https://doi.org/10.1021/jp212541y} {\bibfield
  {journal} {\bibinfo  {journal} {The Journal of Physical Chemistry B}\
  }\textbf {\bibinfo {volume} {116}},\ \bibinfo {pages} {8494} (\bibinfo {year}
  {2012})}\BibitemShut {NoStop}%
\bibitem [{\citenamefont {Miyazawa}\ and\ \citenamefont
  {Jernigan}(1985)}]{Miyazawa1985}%
  \BibitemOpen
  \bibfield  {author} {\bibinfo {author} {\bibfnamefont {S.}~\bibnamefont
  {Miyazawa}}\ and\ \bibinfo {author} {\bibfnamefont {R.~L.}\ \bibnamefont
  {Jernigan}},\ }\bibfield  {title} {\bibinfo {title} {Estimation of effective
  interresidue contact energies from protein crystal structures: quasi-chemical
  approximation},\ }\href {https://doi.org/10.1021/ma00145a039} {\bibfield
  {journal} {\bibinfo  {journal} {Macromolecules}\ }\textbf {\bibinfo {volume}
  {18}},\ \bibinfo {pages} {534} (\bibinfo {year} {1985})}\BibitemShut
  {NoStop}%
\bibitem [{\citenamefont {Woelke}\ \emph {et~al.}(2011)\citenamefont {Woelke},
  \citenamefont {von Eichborn}, \citenamefont {Murgueitio}, \citenamefont
  {Worth}, \citenamefont {Castiglione},\ and\ \citenamefont
  {Preissner}}]{Woelke2011}%
  \BibitemOpen
  \bibfield  {author} {\bibinfo {author} {\bibfnamefont {A.~L.}\ \bibnamefont
  {Woelke}}, \bibinfo {author} {\bibfnamefont {J.}~\bibnamefont {von
  Eichborn}}, \bibinfo {author} {\bibfnamefont {M.~S.}\ \bibnamefont
  {Murgueitio}}, \bibinfo {author} {\bibfnamefont {C.~L.}\ \bibnamefont
  {Worth}}, \bibinfo {author} {\bibfnamefont {F.}~\bibnamefont {Castiglione}},\
  and\ \bibinfo {author} {\bibfnamefont {R.}~\bibnamefont {Preissner}},\
  }\bibfield  {title} {\bibinfo {title} {Development of immune-specific
  interaction potentials and their application in the multi-agent-system
  vaccimm},\ }\href@noop {} {\bibfield  {journal} {\bibinfo  {journal} {PloS
  one}\ }\textbf {\bibinfo {volume} {6}},\ \bibinfo {pages} {e23257} (\bibinfo
  {year} {2011})}\BibitemShut {NoStop}%
\bibitem [{\citenamefont {Chowell}\ \emph {et~al.}(2015)\citenamefont
  {Chowell}, \citenamefont {Krishna}, \citenamefont {Becker}, \citenamefont
  {Cocita}, \citenamefont {Shu}, \citenamefont {Tan}, \citenamefont
  {Greenberg}, \citenamefont {Klavinskis}, \citenamefont {Blattman},\ and\
  \citenamefont {Anderson}}]{Chowell2015}%
  \BibitemOpen
  \bibfield  {author} {\bibinfo {author} {\bibfnamefont {D.}~\bibnamefont
  {Chowell}}, \bibinfo {author} {\bibfnamefont {S.}~\bibnamefont {Krishna}},
  \bibinfo {author} {\bibfnamefont {P.~D.}\ \bibnamefont {Becker}}, \bibinfo
  {author} {\bibfnamefont {C.}~\bibnamefont {Cocita}}, \bibinfo {author}
  {\bibfnamefont {J.}~\bibnamefont {Shu}}, \bibinfo {author} {\bibfnamefont
  {X.}~\bibnamefont {Tan}}, \bibinfo {author} {\bibfnamefont {P.~D.}\
  \bibnamefont {Greenberg}}, \bibinfo {author} {\bibfnamefont {L.~S.}\
  \bibnamefont {Klavinskis}}, \bibinfo {author} {\bibfnamefont {J.~N.}\
  \bibnamefont {Blattman}},\ and\ \bibinfo {author} {\bibfnamefont {K.~S.}\
  \bibnamefont {Anderson}},\ }\bibfield  {title} {\bibinfo {title} {Tcr contact
  residue hydrophobicity is a hallmark of immunogenic cd8+ t cell epitopes},\
  }\href@noop {} {\bibfield  {journal} {\bibinfo  {journal} {Proceedings of the
  National Academy of Sciences}\ }\textbf {\bibinfo {volume} {112}},\ \bibinfo
  {pages} {E1754} (\bibinfo {year} {2015})}\BibitemShut {NoStop}%
\bibitem [{\citenamefont {Shang}\ \emph {et~al.}(2009)\citenamefont {Shang},
  \citenamefont {Wang}, \citenamefont {Niu}, \citenamefont {Meng},
  \citenamefont {Fu}, \citenamefont {Ni}, \citenamefont {Lin}, \citenamefont
  {Yang}, \citenamefont {Chen},\ and\ \citenamefont {Wu}}]{Shang2009}%
  \BibitemOpen
  \bibfield  {author} {\bibinfo {author} {\bibfnamefont {X.}~\bibnamefont
  {Shang}}, \bibinfo {author} {\bibfnamefont {L.}~\bibnamefont {Wang}},
  \bibinfo {author} {\bibfnamefont {W.}~\bibnamefont {Niu}}, \bibinfo {author}
  {\bibfnamefont {G.}~\bibnamefont {Meng}}, \bibinfo {author} {\bibfnamefont
  {X.}~\bibnamefont {Fu}}, \bibinfo {author} {\bibfnamefont {B.}~\bibnamefont
  {Ni}}, \bibinfo {author} {\bibfnamefont {Z.}~\bibnamefont {Lin}}, \bibinfo
  {author} {\bibfnamefont {Z.}~\bibnamefont {Yang}}, \bibinfo {author}
  {\bibfnamefont {X.}~\bibnamefont {Chen}},\ and\ \bibinfo {author}
  {\bibfnamefont {Y.}~\bibnamefont {Wu}},\ }\bibfield  {title} {\bibinfo
  {title} {Rational optimization of tumor epitopes using in silico
  analysis-assisted substitution of tcr contact residues},\ }\href@noop {}
  {\bibfield  {journal} {\bibinfo  {journal} {European journal of immunology}\
  }\textbf {\bibinfo {volume} {39}},\ \bibinfo {pages} {2248} (\bibinfo {year}
  {2009})}\BibitemShut {NoStop}%
\bibitem [{\citenamefont {Miyazawa}\ and\ \citenamefont
  {Jernigan}(1996)}]{Miyazawa1996}%
  \BibitemOpen
  \bibfield  {author} {\bibinfo {author} {\bibfnamefont {S.}~\bibnamefont
  {Miyazawa}}\ and\ \bibinfo {author} {\bibfnamefont {R.~L.}\ \bibnamefont
  {Jernigan}},\ }\bibfield  {title} {\bibinfo {title} {Residue-residue
  potentials with a favorable contact pair term and an unfavorable high packing
  density term, for simulation and threading},\ }\href
  {https://doi.org/10.1006/jmbi.1996.0114} {\bibfield  {journal} {\bibinfo
  {journal} {Journal of molecular biology}\ }\textbf {\bibinfo {volume}
  {256}},\ \bibinfo {pages} {623} (\bibinfo {year} {1996})}\BibitemShut
  {NoStop}%
\bibitem [{\citenamefont {Braun}\ \emph {et~al.}(2020)\citenamefont {Braun},
  \citenamefont {Loyal}, \citenamefont {Frentsch}, \citenamefont {Wendisch},
  \citenamefont {Georg}, \citenamefont {Kurth}, \citenamefont {Hippenstiel},
  \citenamefont {Dingeldey}, \citenamefont {Kruse}, \citenamefont {Fauchere}
  \emph {et~al.}}]{braun2020}%
  \BibitemOpen
  \bibfield  {author} {\bibinfo {author} {\bibfnamefont {J.}~\bibnamefont
  {Braun}}, \bibinfo {author} {\bibfnamefont {L.}~\bibnamefont {Loyal}},
  \bibinfo {author} {\bibfnamefont {M.}~\bibnamefont {Frentsch}}, \bibinfo
  {author} {\bibfnamefont {D.}~\bibnamefont {Wendisch}}, \bibinfo {author}
  {\bibfnamefont {P.}~\bibnamefont {Georg}}, \bibinfo {author} {\bibfnamefont
  {F.}~\bibnamefont {Kurth}}, \bibinfo {author} {\bibfnamefont
  {S.}~\bibnamefont {Hippenstiel}}, \bibinfo {author} {\bibfnamefont
  {M.}~\bibnamefont {Dingeldey}}, \bibinfo {author} {\bibfnamefont
  {B.}~\bibnamefont {Kruse}}, \bibinfo {author} {\bibfnamefont
  {F.}~\bibnamefont {Fauchere}}, \emph {et~al.},\ }\bibfield  {title} {\bibinfo
  {title} {Sars-cov-2-reactive t cells in healthy donors and patients with
  covid-19},\ }\href@noop {} {\bibfield  {journal} {\bibinfo  {journal}
  {Nature}\ }\textbf {\bibinfo {volume} {587}},\ \bibinfo {pages} {270}
  (\bibinfo {year} {2020})}\BibitemShut {NoStop}%
\bibitem [{\citenamefont {Murugan}\ \emph {et~al.}(2012)\citenamefont
  {Murugan}, \citenamefont {Mora}, \citenamefont {Walczak},\ and\ \citenamefont
  {Callan}}]{murugan2012}%
  \BibitemOpen
  \bibfield  {author} {\bibinfo {author} {\bibfnamefont {A.}~\bibnamefont
  {Murugan}}, \bibinfo {author} {\bibfnamefont {T.}~\bibnamefont {Mora}},
  \bibinfo {author} {\bibfnamefont {A.~M.}\ \bibnamefont {Walczak}},\ and\
  \bibinfo {author} {\bibfnamefont {C.~G.}\ \bibnamefont {Callan}},\ }\bibfield
   {title} {\bibinfo {title} {Statistical inference of the generation
  probability of t-cell receptors from sequence repertoires},\ }\href
  {https://doi.org/10.1073/pnas.1212755109} {\bibfield  {journal} {\bibinfo
  {journal} {Proceedings of the National Academy of Sciences}\ }\textbf
  {\bibinfo {volume} {109}},\ \bibinfo {pages} {16161} (\bibinfo {year}
  {2012})}\BibitemShut {NoStop}%
\end{thebibliography}%

\end{document}